\documentclass[fleqn,usenatbib]{mnras}
\pdfoutput=1

\DeclareRobustCommand{\VAN}[3]{#2}
\let\VANthebibliography\thebibliography
\def\thebibliography{\DeclareRobustCommand{\VAN}[3]{##3}\VANthebibliography}

\usepackage{graphicx}	% Including figure files
\usepackage{amsmath}	% Advanced maths commands
\usepackage{newtxtext,newtxmath}
\usepackage[T1]{fontenc}

\usepackage{xcolor}

\title[Morphological classification of galaxies]{Morphological classification of galaxies with deep learning: comparing 3-way and 4-way CNNs}

\author[M. K. Cavanagh et al.]{
Mitchell K. Cavanagh,$^{1}$\thanks{E-mail: mitchell.cavanagh@icrar.org (MKC)}
Kenji Bekki$^{1}$
and Brent A. Groves$^{1,2}$
\\
$^{1}$ICRAR M468, The University of Western Australia, 35 Stirling Hwy, Crawley, WA 6009, Australia,\\
$^{2}$Research School of Astronomy and Astrophysics (RSAA), Australian National University, ACT 2611, Australia
}

\date{Accepted XXX. Received YYY; in original form ZZZ}

\pubyear{2021}

\begin{document}
\label{firstpage}
\pagerange{\pageref{firstpage}--\pageref{lastpage}}
\maketitle

\begin{abstract}
Classifying the morphologies of galaxies is an important step in understanding their physical properties and evolutionary histories. The advent of large-scale surveys has hastened the need to develop techniques for automated morphological classification. We train and test several convolutional neural network architectures to classify the morphologies of galaxies in both a 3-class (elliptical, lenticular, spiral) and 4-class (+irregular/miscellaneous) schema with a dataset of 14034 visually-classified SDSS images. We develop a new CNN architecture that outperforms existing models in both 3 and 4-way classification, with overall classification accuracies of 83\% and 81\% respectively. We also compare the accuracies of 2-way / binary classifications between all four classes, showing that ellipticals and spirals are most easily distinguished (>98\% accuracy), while spirals and irregulars are hardest to differentiate (78\% accuracy). Through an analysis of all classified samples, we find tentative evidence that misclassifications are physically meaningful, with lenticulars misclassified as ellipticals tending to be more massive, among other trends. We further combine our binary CNN classifiers to perform a hierarchical classification of samples, obtaining comparable accuracies (81\%) to the direct 3-class CNN, but considerably worse accuracies in the 4-way case (65\%). As an additional verification, we apply our networks to a small sample of Galaxy Zoo images, obtaining accuracies of 92\%, 82\% and 77\% for the binary, 3-way and 4-way classifications respectively.
\end{abstract}


%%%%%%%%%%%%%%%%%%%%%%%%%%%%%%%%%%%%%%%%%%%%%%%%%%

%%%%%%%%%%%%%%%%% BODY OF PAPER %%%%%%%%%%%%%%%%%%

\section{Introduction}

Galaxies exhibit a variety of morphologies throughout cosmic time \citep{Conselice2014}, ranging from broad families of spirals to large ellipticals. A galaxy's morphology is related to its internal physical processes \citep{Sellwood2014}, such as its dynamical and chemical evolution, star formation activity \citep{Kennicutt1998}, and can reveal insights into its past evolutionary history, including mergers \citep{Mihos1996} and interactions with its environment \citep{SolAlonso2006}. A galaxy's morphology is often determined by visual inspection, and while this is sufficient to delineate a rich array of morphological types in the nearby Universe \citep{Buta2013}, there is the inherent issue of scalability as future surveys yield far greater magnitudes of data.

Visually classifying the apparent morphologies of galaxies through the manual inspection of images is a laborious, time-consuming task. Recently, the use of citizen science has seen great success in classifying larger datasets \citep{Lintott2011, Willett2013, Simmons2016}, however it is important to keep in mind that this was achieved by increasing the order of magnitude of human classifiers, while the speed of the actual classification itself is more or less unchanged. Future large-scale surveys are expected to yield enormous amounts of data, with \textit{Euclid} alone expected to survey at least to the order of 1 billion galaxies. The scale of surveys such as these will overwhelm the capacity of human volunteers \citep{Silva2018}, necessitating the development and deployment of techniques for automated classification.

Automated classification techniques, in particular those utilising neural networks \citep{Dieleman2015,DomingeuzSanchez2018,Zhu2019}, have the ability to revolutionise the speed at which samples can be individually classified. Such classification techniques range from analytical approaches, such as the Fourier analysis of luminosity profiles \citep{Odewahn2002} and isophote decompositions \citep{MndezAbreu2017}, to statistical learning methods \citep{Sreejith2017}, random forest classifiers \citep{Beck2018} and ensemble classifiers \citep{Baqui2021}, which are all part of a broader suite of machine learning methods \citep{Cheng2020}. Neural networks have been utilised in this field for some time, but only recently has the rapidly growing field of deep learning seen widespread applications in astronomy \citep{Baron2019}. One of the key architectures behind the success of deep learning, particularly in applications involving image recognition and computer vision, is the convolutional neural network (hereafter CNN), introduced by \citet{Lecun1998}. Unlike the typical multiplayer perceptrons present in a normal neural network, CNNs are specifically designed to extract features in data through multiple convolutions \citep{LeCun2015}. Each convolution can be thought of as a layer of abstraction, with the extracted high-level features used to learn a generalised representation of data. Importantly, CNNs provide a model-independent means of feature extraction, hence their versatile range of applications. Although having originally been developed to recognise handwritten characters, modern CNNs are capable of general image recognition across myriad image types \citep{He2015}. Indeed, image classification is one of the main astronomical applications of CNNs \citep{Cheng2020}.

CNNs have been successfully utilised to detect quasars and gravitational lenses \citep{PasquetItam2018, Schaefer2018}, study bulge/disk dominance \citep{Ghosh2020}, detect stellar bars \citep{Abraham2018} and classify radio morphologies \citep{Wu2018}. CNNs have also been widely used with simulations, including cosmological simulations \citep{Mustafa2019} and mock surveys \citep{Ntampaka2020}, as well as to develop tools for galaxy photometric profile analysis \citep{Tuccillo2017}. Recent studies have utilised CNNs for the purpose of binary classification \citep{Ghosh2020}, or classifying between general morphological shapes \citep{Zhu2019}. Fewer studies have looked at 3-way classification between distinct morphological types, though some works have explored classifying between ellipticals and barred/unbarred spirals \citep{Barchi2020}, or between ellipticals, spirals and irregulars \citep{DeLaCalleja2004}

In this work, we train and test several different CNN architectures for the purpose of initially distinguishing between elliptical (E), lenticular (S0) and spiral (Sp) galaxies, before extending the schema to include a fourth irregular and miscellaneous category (Irr+Misc). We base our training data on the visually classified  catalogue of \citet{Nair2010}, consisting of 14034 samples from the SDSS Data Release 4 \citep{AdelmanMcCarthy2006}, and apply a series of data augmentation techniques to increase the number of training samples for use with the CNN. Our best accuracies are consistent with previous studies, surpassing others when it comes to per-class accuracies. We dedicate a section of our discussion to focusing on the physical implications of the CNN classifications (both correct and incorrect), as well as how these are reflective of the inherent uncertainty in the training data, how such uncertainties affect CNN accuracies, and how CNNs can be used to address these uncertainties.

The structure of the paper is as follows. In \S 2 we briefly outline the theory of CNNs and its key concepts and terminology, discuss the CNN architectures including our new C2 network, discuss the augmentation and training methodology, and discuss the overall performance of the 3-way and 4-way classification tasks for each of the four CNN architectures. In \S 3 we present our key results, starting with binary classifications between morphological classes, before presenting the best results for the 3-way and 4-way classifications and analysing these by considering the physical properties of the samples for each classification. In \S 4 we discuss the accuracy of the CNN. We examine the physical properties of classified samples and show that there are common trends in the misclassified samples across several physical properties. We showcase binary hierarchical classification as an alternative to the direct multi-class CNNs, exploring five different classifiers and their key differences to our main CNN. We present several examples of images that were correctly and incorrectly classified, commenting on inherent uncertainties and the challenge of unambiguously classifying samples, whether by eye or by CNN. As a final, independent verification of our network, we apply our models to a small sample consisting of 1,000 randomly chosen ellipticals and 1,000 randomly chosen spiral galaxies from the Galaxy Zoo dataset. Lastly, we summarise our findings in \S 5.

\section{Methods}

\subsection{Dataset}

The \citet{Nair2010} catalogue, hereafter NA10, contains 14,304 visually classified samples from the SDSS Data Release 4 \citep{AdelmanMcCarthy2006}. These are all single-band, g-band images with spatial resolutions of 50 kpc. Of key interest to this work are the Hubble T-Type classifications, which range from -5 to 12 with a miscellaneous 99 category for samples with no clear category. The miscellaneous category also includes interacting galaxies and mergers. One of the goals of the catalogue is to serve as a dataset with which to calibrate automated classification techniques.

\begin{table*}
\centering
\caption{Definition of our four morphological categories in terms of Hubble morphologies and numerical T-Types schemas. Ellipticals are defined as c0 to E+ (Nair T-Type -5 only), S0s defined as S0- to S0/a (Nair T-Types -3 to 0), Spirals as Sa to Sm (Nair T-Types 1 to 9), and Irr+Misc as Im onwards (Nair T-Types 10 and 99).}
\label{tab:ttypes}
\begin{tabular}{|c|c|c|c|c|c|c|c|c|c|c|c|c|c|c|c|c|c|c|}
\hline 
Category & \multicolumn{3}{|c|}{Elliptical} & \multicolumn{4}{|c|}{S0} & \multicolumn{9}{|c|}{Spiral} & \multicolumn{2}{|c|}{Irr+Misc} \\ 
\hline
Class & c0 & E0 & E+ & S0- & S0 & S0+ & S0/a & Sa & Sab & Sb & Sbc & Sc & Scd & Sd & Sdm & Sm & Im & Misc \\ 
RC3 & -6 & -5 & -4 & -3 & -2 & -1 & 0 & 1 & 2 & 3 & 4 & 5 & 6 & 7 & 8 & 9 & 10 & - \\  
Nair & -5 & -5 & -5 & -3 & -2 & -2 & 0 & 1 & 2 & 3 & 4 & 5 & 6 & 7 & 8 & 9 & 10 & 99 \\ 
\hline 
\end{tabular} 
\end{table*}

For this work, we consider up to four morphological categories; ellipticals (E), lenticulars (S0), spirals (Sp) and irregulars plus miscellaneous types (Irr+Misc). The samples in the Nair catalogue have T-Types that range from -5 to 12, with a miscellaneous T-Type 99. Table \ref{tab:ttypes} outlines our definition of the four morphological classes, along with the NA10 T-Types and the classic RC3 \citep{deVaucouleurs1991} schema for reference. Using these definitions, 2723 samples are categorised as ellipticals, 3215 as S0s, 7708 as Spirals and 388 as Irr+Misc.

\subsection{Convolutional Neural Networks}

In this section we briefly outline the theory of neural networks, introduce some key terminology and explore how CNNs differ from standard neural networks. A standard, fully-connected, feed-forward artificial neural network consists of multiple layers of nodes, such that each node is connected to all the nodes in its preceding and succeeding layers \citep{Haykin2009, Goodfellow2016}. The input data is mapped to the nodes in the input layer, with the nodes in the final layer corresponding to the outputs. Each node has an activation value $x$ that is obtained as follows:
\[ x_i^l = \phi\left(\sum_{j}^{N_{l-1}} x_j^{l-1} w_{i,j}^{l} + b_i^l\right) \]
where $l$ denotes the current layer, $x_i^l$ denotes the current node, $N_{l-1}$ is the number of nodes in the previous layer, $x_j^{l-1}$ is the activation of the $j$th node in the previous layer, $w_{i,j}^l$ is the weight assigned to the edge connecting the current node $x_i^l$ to the previous node $x_j^{l-1}$, $b_i^l$ is a bias, and $\phi$ is an activation function. To train a neural network, we supply it with at least two similarly representative sets of data; a training set, and a test set.
A third validation set, independent of the test set, is used to tune the hyperparameters of the network.
The neural network is trained by adjusting the weights $w^l$ and biases $b^l$ for all nodes in every layer $l$ to minimise the error (with respect to an objective or loss function) with the desired outputs in the training set. This is usually achieved via a learning algorithm such as gradient descent over multiple iterations of training, where each iteration is referred to as an epoch. Once trained over a sufficiently large number of epochs, the final success rate is determined via an evaluation on an unseen test set. The training is typically conducted in batches, with the weights and biases updated for each batch.
 
CNNs are a specialised type of neural network designed to extract high-level features from the provided input data through the application of convolutions. Each 2D convolutional layer takes an $s \times s$ input array and applies $m$ filters/weights using a $k \times k$ kernel (a.k.a window), where $k < s$, in order to obtain $m$ unique feature maps. In particular, if the input to a convolutional layer is a set of $N$ matrices $\mathbf{X}_{l-1}^{n}$ for $n = 1, \ldots, N$, then the feature maps are given as
\[ \mathbf{X}_l^{(m)} = f\left(\sum_{n=1}^N \mathbf{W}_l^{(n,m)} * \mathbf{X}_{l-1}^{(n)} + b_l^{(n)} \right) \]  
where $m$ denotes the $m$th feature map, $\mathbf{W}_l^{(n,m)}$ are the filters/weights for layer $l$, input map $n$ and feature map $m$, $b_l^{(n)}$ is a vector of biases, and $f$ is a (multi-dimensional) activation function. Here $*$ denotes a linear convolution. Note that the inputs and feature maps need not be square. Importantly, each feature map is convolved from the input image with a different set of filters/weights. These weights $\mathbf{W}_l^{(n,m)}$ are learnable and are tweaked during training. Key to the success of CNNs is that they learn multiple filters in parallel, which enables the network to not only obtain different abstract representations of the data, but also use these in order to distinguish between data of different types.

Unlike in a normal neural network, where each node's edges are free parameters, feature maps use shared weights by applying a single convolutional filter across the entire input, rather than separate filters for individual parts of the input. Thus, in order to minimise the loss with the training data, the CNN directly adjusts these filter weights, hence obtaining different feature maps. As such, CNNs provide a model independent means to classify images, relying solely on self-taught convolution weights to extract the most meaningful features in data. Successive convolutional layers are used to extract higher-level features. Generally, later convolutional layers extract more feature maps with smaller kernels.

CNNs also contain another important layer known as a pooling layer. These layers are used to down sample feature maps \citep{Zhou2015}. For this work, we use MaxPooling. This method of pooling preserves the maximal elements, helping to highlight dominant features. This also helps to introduce a degree of translational invariance and also serves to reduce noise. After the features have been extracted via convolutional and pooling layers, all the feature maps are flattened into a single 1D array. These are then fully connected to subsequent layers of nodes (referred to as Dense layers), ultimately ending with the output layer which contains a total number of nodes equal to the total number of output categories. These layers are functionally identical to a normal neural network. After flattening the feature maps, it is common to remove a fraction of randomly selected nodes. This is an efficient regularisation technique known as Dropout, and is used to reduce overfitting of the data and improve generalisation \citep{Srivastava2014}.

\subsection{Model Architectures}

\begin{figure*}
\centering
\includegraphics[scale=0.18]{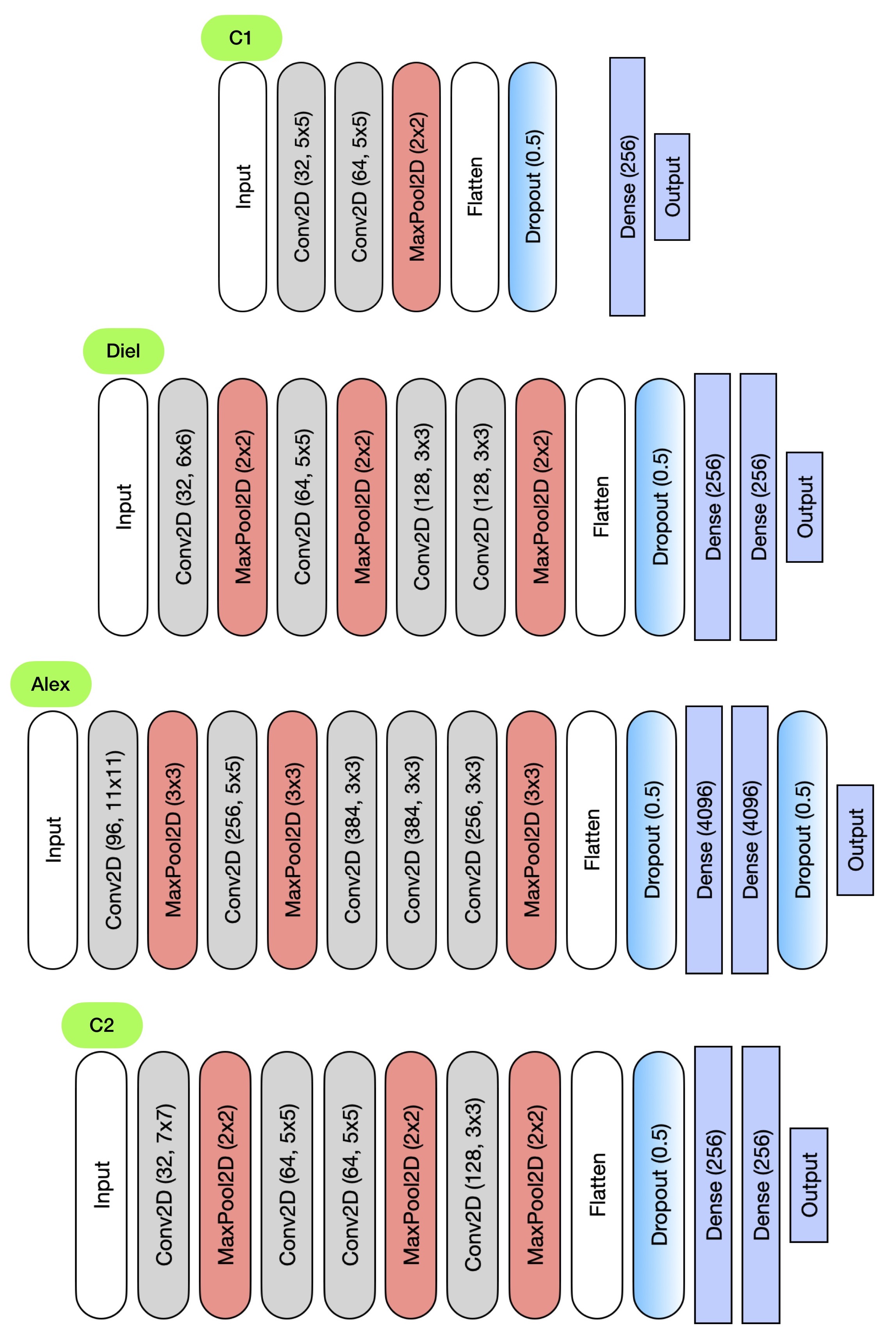}
\caption{Schematic representation of all four principal CNN architectures used in this work: the CNN from \citet{Cavanagh2020} (C1), the CNN based on \citet{Dieleman2015} (Diel), the CNN based on \citet{Krizhevsky2012} (Alex), and our new CNN introduced in this work (C2). Layers are colour coded: grey for convolutional layers, red for pooling, light blue for dropout and blue for dense layers. Convolutional layers are annotated with the number of feature maps, followed by the kernel size. Pooling layers are annotated with the size of the pooling. Dropout layers are annotated with the dropout rate (the fraction of nodes ignored in the previous layer). Dense layers are annotated with the number of nodes.}
\label{fig:architectures}
\end{figure*}

In this work we consider four principal CNN architectures, each with different complexity and originally designed for different input image sizes. All architectures are designed to accept single-band images as their input. Figure \ref{fig:architectures} displays a schematic illustration of each architecture. Each architecture is built using Keras \citep{Chollet2015}, a high-level machine learning interface. The first CNN, referred to as C1, is based on our previous work \citep{Cavanagh2020}, and was originally designed for the binary classification of $50 \times 50$ imagery. This architecture consists of a single block of convolution and pooling layers (2 Conv2D plus a MaxPooling), a penultimate Dense layer with 256 nodes, and the output Dense layer which contains either 3 or 4 nodes for 3-way or 4-way classification.

The second CNN architecture is based on the work of \citet{Dieleman2015}. This architecture contains multiple convolutional blocks, starting with two blocks of alternating, single Conv2D and MaxPool layers, then a block of two successive Conv2D layers followed by a final MaxPool. There are also two Dense layers before the output layer. Key differences to the original implementation in \citet{Dieleman2015} is that we make use of the Adadelta adaptive learning rate algorithm for training \citep{Zeiler2012}, and we have also added Batch Normalisation layers at the end of every convolutional block. When training a network, the training data is usually processed in batches, with the weights and biases tweaked after each batch. Batch normalisation is used to enforce uniformity across all samples in a batch by keeping a consistent mean and standard deviation. This has the effect of smoothing the objective function (hence smoother gradients), leading to better performance, improved regularisation and generalisation \citep{Ioffe2015,Santurkar2019}. 

The third architecture is based on AlexNet \citep{Krizhevsky2012}, a network designed for general image classification. This architceture has commonly been used in many classification problems, including most recently in astronomy \citep{Ghosh2020}. This is the most complex architecture out of the four, and is notable for its heavy use of dropout regularisation with two Droput layers, each dropping 50\% of the previous layer's nodes. The convolutional layers also utilise a large number of filters. The original optimiser used for the training of this network is stochastic gradient descent with momentum, however the learning rate is computationally expensive to tune. Instead we use the Adadelta adaptive learning rate algorithm. We also add batch normalisation layers after each convolutional block but before the max pooling, adding even more regularisation to the network.

The fourth architecture, named C2, is completely new from this work, designed as a three-block CNN with two dense layers, a single dropout layer, and batch normalisation throughout. Unlike previous models, we decided to use the Adam (Adaptive Moment Estimation) algorithm \citep{Kingma2014}, an efficient stochastic gradient descent algorithm that can be considered an extension of Adadelta. Unlike Adadelta, Adam's learning rate must be pre-specified. To determine the optimal design for our C2 architecture, including the best learning rate, we performed hyperparameter tuning with the tool Keras Tuner \citep{Omalley2019}. A hyperparameter is any value that affects the architecture of a neural network (e.g. number of nodes) or the manner in which it is trained (e.g. learning rate). Hyperparameter tuning refers to the process of optimising the hyperparameters within the search space as specified by the range of desired values. In the case of C2, we started with the same sequence of layers in Figure \ref{fig:architectures}, and used hyperparameter tuning to determine the optimal number of feature maps for each convolutional layer (out of a choice of 32 or 64 for the first convolutional layer, and 64 or 128 for all layers thereafter), width of the convolutional filter or kernel (7x7 or 5x5 for the first layer, 5x5 or 3x3 for all others), number of nodes in the dense layers (256, 512 or 1024), degree of dropout (0.3, 0.4 or 0.5), whether to use max pooling or average pooling, and finally the learning rate for the Adam optimiser, to be determined by sampling between $10^{-3}$ and $10^{-4}$. The optimal parameters for the C2 architecture, as determined by hyperparameter tuning, are annotated in Figure \ref{fig:architectures}. The final, tuned learning rate for the Adam optimiser was $2 \times 10^{-4}$.

Across all architectures, we used consistent activation functions, with ReLU for all convolutional and dense layers, and with a final softmax activation for the output layer. Softmax, a type of normalised exponential, is important as it converts the activations into probabilities that sum to 1 across all output categories. The predicted category, or simply the prediction, is just the category with the maximum value (in other words, the output node with the highest activation). This maximum value itself is often referred to as the probability or confidence; we will use both terms interchangeably.

\begin{figure*}
\centering
\includegraphics[scale=0.65]{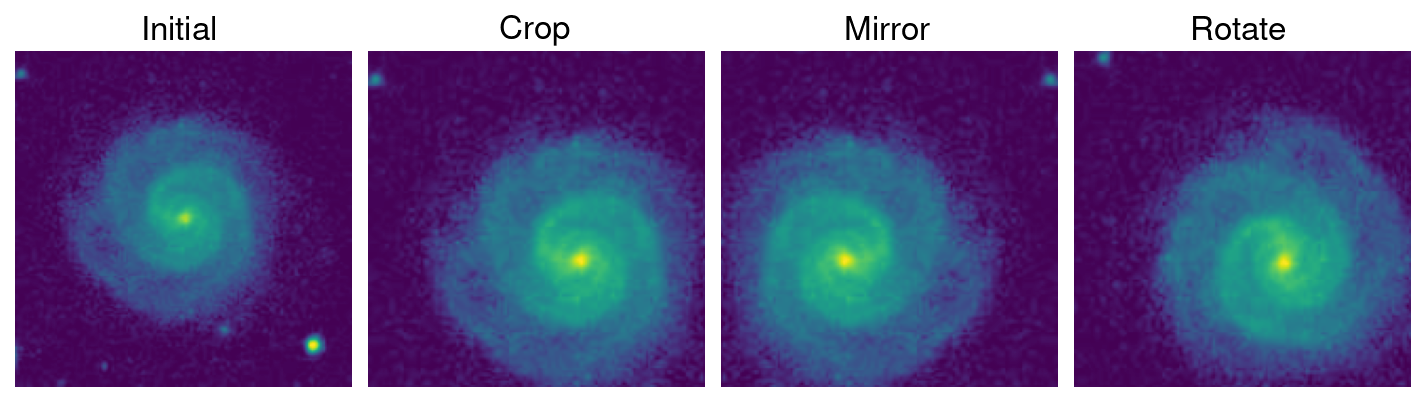}
\caption{An example of our principal data augmentation techniques (cropping, mirroring and rotating) as applied to a spiral galaxy. The original image is on the left.}
\label{fig:augexample}
\end{figure*}

\subsection{Data Augmentation and Pre-Processing}

The size of the NA10 dataset is relatively small given the complexity of the CNN architectures, and is thus prone to overfitting. In order to increase the size of the dataset, and subsequently improve the performance of the CNN, we apply several data augmentation techniques. These are listed as follows:
\begin{itemize}
\item Cropping. For an image of size $n \times n$ we can crop multiple $m \times m$ images for $n > m$. We have chosen to crop each image from each corner, as well as in the center, for a total of five cropped images. Cropping was done with an $110 \times 110$ image to yield five $100 \times 100$ images.
\item Rotation. We rotate each image by 90$^\circ$ increments, yielding four images.
\item Flipping / Mirroring. Yields two new images. Our implementation is equivalent to a flip about the axis $y = x$.
\end{itemize}
As implemented, these techniques combine to increase the number of images by a factor of 40, from 14,034 to a maximum of 561,360 images. An example of these techniques as applied to a sample spiral galaxy from our training set is shown in Figure \ref{fig:augexample}. These data augmentation techniques also help to generalise the input data and enforce a degree of spatial and translational invariance. This is important, as the CNN must learn to identify the features themselves, and not simply that they are present in specific regions within the feature maps. Based on preliminary experiments with a small subset of the total data, we found that both rotating and flipping had a noticeable, positive impact on the overall training accuracy, albeit similar effects due to cropping were not as pronounced. We chose only to process five cropped images to maintain a reasonable file size while keeping within our computational limitations; any further cropping is a case of diminishing returns given the lack of any meaningful improvement to the overall accuracy.

In this study, our primary resolution for the input image data is $100 \times 100$. We also perform some further runs at $200 \times 200$ to assess the impact of changing resolution, however it is important to note that the prohibitively expensive memory and computational requirements for the $200 \times 200$ runs restricted the degree of augmentation (in particular, no cropping). This is since the limited amount of GPU memory available necessarily places an upper limit on the total number of samples that can be trained with at a given resolution. Instead, we direct most of our focus to the fully augmented $100 \times 100$ data. The original imagery in the NA10 dataset comes in variable resolutions, the most common of which were between 100 and 130 pixels. These were all resized to the desired, target resolution using cubic interpolation with Pillow, a fork of the Python Imaging Data. The raw pixel data was extracted and reshaped into the correct format for loading into the CNN. All input images were linearly normalised such that pixel values are between 0 and 1. No further alterations were made to the images.

\subsection{Training and Validation}

\begin{figure*}
\centering
\includegraphics[scale=0.75]{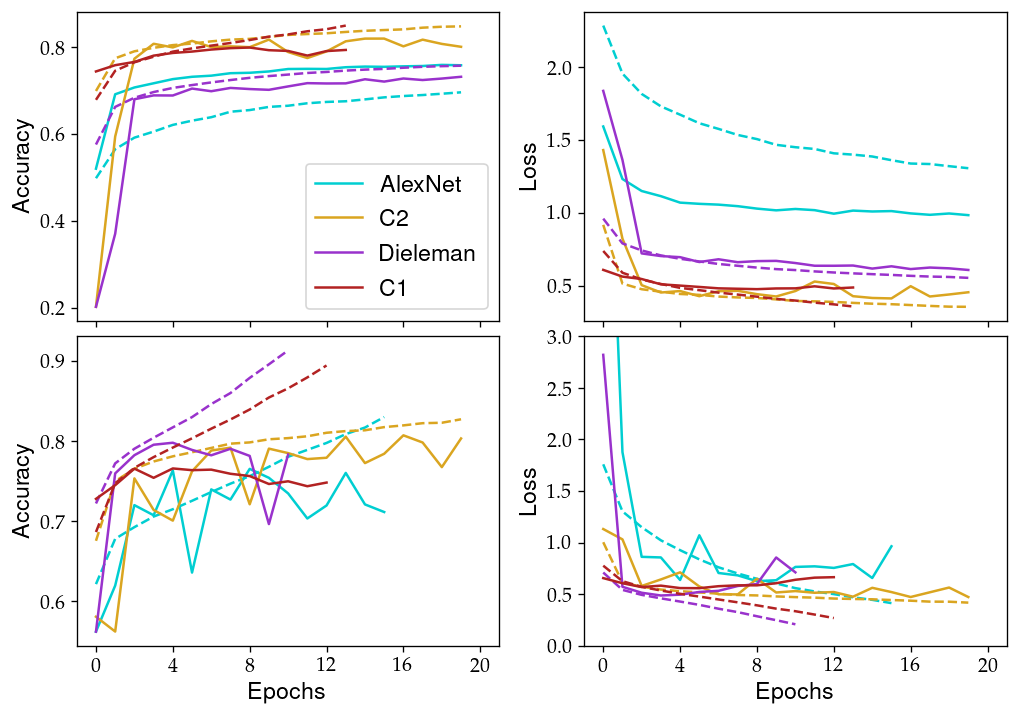}
\caption{Accuracies and loss for all four CNN architectures for the 3-way (top panels) and 4-way (bottom panels) classification schemas as functions of the number of epochs, and as trained on the $100 \times 100$ data. The training accuracy/loss is represented with a dotted line, while the validation accuracy/loss is represented with a solid line. Architectures are represented with C1 in red, C2 in yellow, AlexNet in cyan and Dieleman in magenta.}
\label{fig:traintest}
\end{figure*}

The CNNs were constructed with the high-level API Keras \citep{Chollet2015} and trained using the TensorFlow machine learning library running with nVidia's GPU programming toolkit CUDA. Training was conducted on ICRAR's \textit{Hyades} cluster, making use of a Nvidia GTX 1080 Ti graphics card, with smaller-scale runs and testing performed on a Nvidia GTX 1650 Ti laptop graphics card. The dataset was first partitioned into separate training and test sets according to an 85:15 split, i.e. 15\% of the dataset is reserved for testing, with 85\% used for training. The training set was further partitioned into separate training and validation sets (again with an 85:15 split) for the purpose of hyperparameter tuning. A third validation set, fully independent of the actual test set, is necessary for hyperparameter tuning in order to avoid biasing the choice of hyperparameters. Thus, when the optimal network is finally trained, it is evaluated on an independent, unseen dataset. This is an important test of the network's ability to effectively generalise.

Training was conducted over a maximum of 100 epochs with a batch size of 400 for $100 \times 100$ data, and a reduced batch size of 200 for the $200 \times 200$ data due to memory limitations. To conduct these runs efficiently, we used early stopping; a method of ceasing training once there is no appreciable improvement in the validation loss over a threshold number of epochs. This threshold - the number of epochs over which to monitor whether the training is improving or not - is known as the patience. All runs were conducted with a patience of 7, and had generally converged and been stopped by the 20th to 25th epoch. Early stopping helps to avoid wasting unnecessary computations once the validation accuracy has plateaued, and subsequently prevents overfitting. The best weights were stored using Keras' ModelCheckpoint callback function. We make use of the categorical cross-entropy loss function for all CNNs.

Figure \ref{fig:traintest} shows all training and final validation accuracies/loss for all four CNN architectures. Note here that the validation accuracy/loss is merely a metric based on evaluations on the test set. All the weights and biases are initially randomised at the start of training. It can be seen that the validation accuracies initially rise rapidly before plateauing. Similarly, the validation loss gradually falls before becoming constant; indeed, early stopping is used to halt the training when the loss no longer decreases. Were early stopping not employed, the validation loss would continue to increase (see the C1 and AlexNet 4-way plots in the bottom-right panel of Figure \ref{fig:traintest}), further widening the gap with the training loss and resulting in overfitting. The accuracy and loss values exhibit large random fluctuations, especially in the 4-way case, which is another sign of overfitting. As can be seen in Figure \ref{fig:traintest}, our C2 network is the best overall performer, since it is less affected by overfitting and achieves lower overall validation losses compared to the other models. Note the validation accuracy and loss are purely metrics corresponding to repeated evaluations of the network with the independent test set at the conclusion of every epoch. In Keras, these steps do not affect any of the network parameters.

\begin{figure*}
\centering
\includegraphics[scale=0.35]{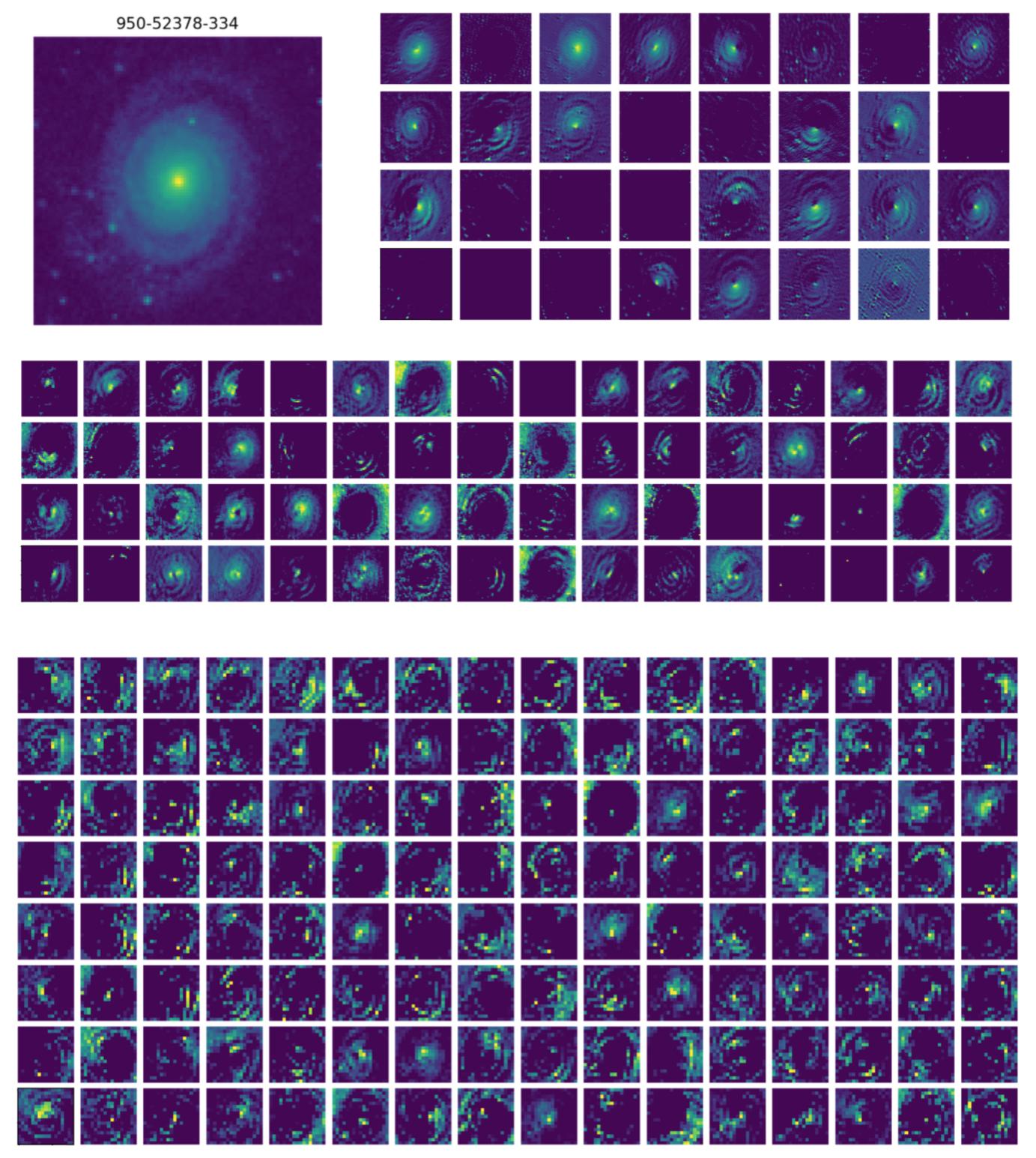}
\caption{Plot of each individual feature map in the first, third and final convolutional layers of the C2 network, as generated for the example spiral galaxy shown in the top-left.}
\label{fig:cnn-vis}
\end{figure*}

It is useful to visualise the feature maps within the convolutional and pooling layers in order to understand how the network is interpreting the input data; in particular, the features that are being extracted via the learnable filter weights. Figure \ref{fig:cnn-vis} shows every individual feature map in the first, third and fourth convolutional layers of the C2 CNN architecture (see Figure \ref{fig:architectures}) for the shown spiral galaxy in the 3-way case. It can be seen that, after more and more convolutions and poolings, the features maps become more abstract. Through the use of these convolutions, the CNN is able to extract relevant features from the input image, such as spiral arms, the central bulge, and even a trace of the outline of the galaxy itself. It is important to stress that the filters/weights for each convolution are initially random and have instead been learned by the network. Thus visualising the feature maps enables us to see the results of these convolutions, and in the case of Figure \ref{fig:cnn-vis}, the images that the CNN has learned to interpret as a spiral galaxy.

\section{Results}

\subsection{Binary classifications}

\begin{figure*}
\centering
\includegraphics[scale=0.65]{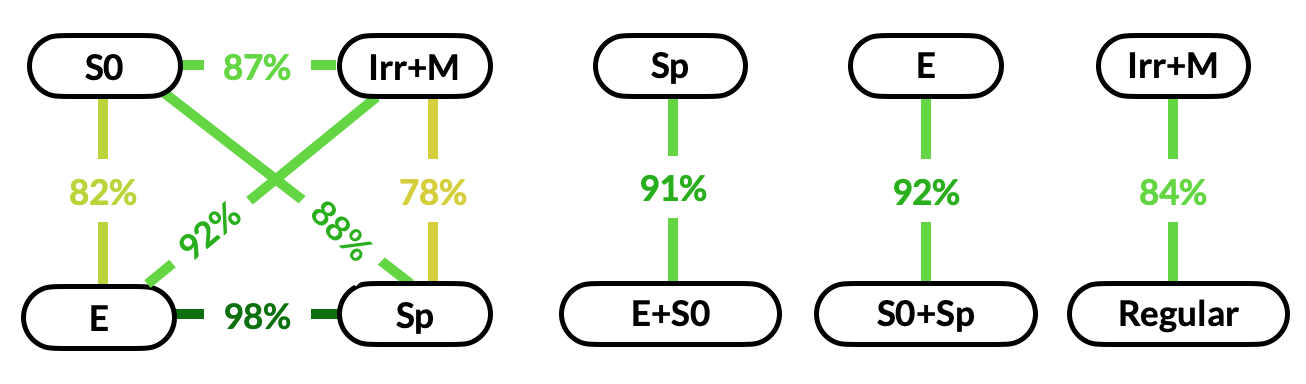}
\caption{Binary classification accuracies for various combinations of morphological types as trained and tested with the C2 network. Accuracies are coloured according to relative accuracy, with dark green corresponding to the strongest distinction (E vs. Sp at 98\%) and olive corresponding to the weakest distinction (Sp vs. Irr+Misc at 78\%).}
\label{fig:hier-acc}
\end{figure*}

Before classifying galaxies as either ellipticals, lenticulars, spirals or irregulars/misc, it is useful to first consider binary classifications between all four of these classes. This allows us to anticipate which distinctions are most problematic and likely to be sources of confusion. This will further serve to assist in interpreting the confusion matrices of the 3-way and 4-way classification.

We conducted two-way classifications between all four morphological classes (E vs S0, E vs Sp, $\ldots$) using our newly developed C2 network architecture. We further conducted two-way classifications between combinations of classes: Sp vs E and S0 (this can be considered an early vs. late type classifier), E vs. S0 and Sp, and irregulars vs. regulars, where regulars are subsampled from all E, S0 and Sp samples. Since there are an extremely low number irregulars in the training set compared to the other classes, the 2-way classifications involving irregulars used a training set with an equal number of samples from each class, hence the over-represented class was randomly subsampled. In these cases, the performance is dependent on the quality of the subsample (insofar as its resemblance, or lack thereof, to the full sample), so each run was repeated multiple times with different subsamples, with the mean accuracy taken as the final result. The ``regulars'' training set used for the irregular vs. regular run was created with an equal number of samples from the E, S0 and Sp categories. All other runs not involving irregulars (e.g. E vs Sp) were conducted with all the available data, hence needing only a single run. All datasets were partitioned according to an 85:15 split into training and testing sets.

Figure \ref{fig:hier-acc} shows the classification accuracies for all the aforementioned binary classifications. Ellipticals and spirals are able to be distinguished with a very high accuracy (>98\%), spirals and lenticulars at 88\%, ellipticals and irregulars at 86\%, spirals and irregulars at 79\%, ellipticals and lencitulars at 78\%, and lastly lenticulars and irregulars at 70\%. We also find that spirals can be disassociated from Es and S0s combined with a high accuracy (91\%); similarly for Es vs. S0s and spirals (92\%). Lastly, irregulars vs. regulars had an accuracy of 84\%, although it is important to keep in mind that this used a considerably smaller training set as there were so few irregulars. These results are consistent with expectations from previous, established studies, e.g. \citet{DeLaCalleja2004} obtained a 95\% accuracy for E vs Sp.

\subsection{3-way and 4-way classifications}

Many previous studies utilising CNNs in astronomy have used binary classifications. In this study, we directly train a CNN to classify firstly between three classes, and again between four. Table \ref{tab:acc2} presents a summary of the overall classification accuracies for all CNN architectures with both $100 \times 100$ and $200 \times 200$ data. Note that memory limitations prevented the C1 architecture being used with the $200 \times 200$ data. Our C2 architecture outperforms all other models with the $100 \times 100$ data, with only the $200 \times 200$ 4-way AlexNet run beating it. We obtain accuracies of 83\% and 81\% for 3-way and 4-way classifications respectively with $100 \times 100$ data. Increasing the resolution to $200 \times 200$ has a marginal effect on the overall classification accuracies, with only the AlexNet architecture producing a consistent gain. Given the resultant four-fold increase in computationally complexity and memory requirements for little comparative gain, we instead base our analysis on the $100 \times 100$ data. Our achieved accuracies are comparable to recent CNN applications, e.g. \citet{Barchi2020} with 83\% for 3-class classification (albeit for ellipticals and barred/unbarred spirals).

\begin{table}
\centering
\caption{Best classification accuracies of the principal CNN architectures for different input sizes and number of output categories, with 3-way distinguishing between E, S0 and Sp, and 4-way distinguishing between E, S0, Sp and Irr+Misc.}
\label{tab:acc2}
\begin{tabular}{|c|c|c|c|}
\hline 
{\bf Network} & {\bf Input Size} & {\bf Best 3-Way} & {\bf  Best 4-Way} \\ 
\hline 
C1 & $100 \times 100$ & 79\% & 75\% \\ 
\hline 
Dieleman & $100 \times 100$ & 82\% & 76\% \\ 
\hline 
\ & $200 \times 200$ & 82\% & 79\% \\ 
\hline 
AlexNet & $100 \times 100$ & 75\% & 74\% \\ 
\hline 
\ & $200 \times 200$ & 84\% & 83\% \\ 
\hline
C2 & $100 \times 100$ & 83\% & 81\% \\
\hline
\ & $200 \times 200$ & 81\% & 80\% \\
\end{tabular} 
\end{table}

\begin{figure}
\centering
\includegraphics[scale=0.6]{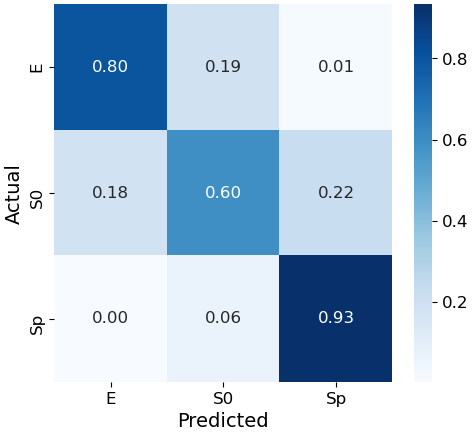}
\caption{The confusion matrix of our C2 network for the 3-way classification between ellipticals (E), lenticulars (S0) and spirals (Sp), coloured according to the accuracy of the classification.}
\label{fig:confmat3way}
\end{figure}

\begin{figure}
\centering
\includegraphics[scale=0.6]{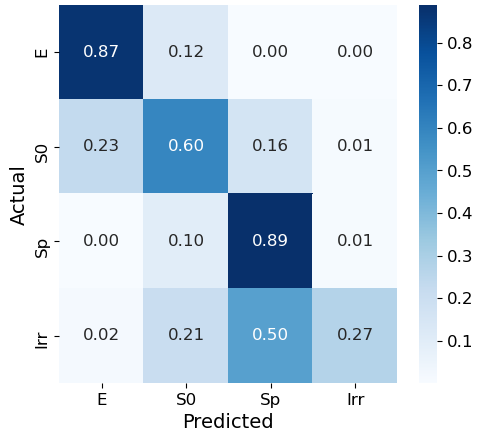}
\caption{The confusion matrix of our C2 network for the 4-way classification between ellipticals (E), lenticulars (S0), spirals (Sp) and irregular/misc samples (Irr), coloured according to the accuracy of the classification.}
\label{fig:confmat4way}
\end{figure}

Accuracy alone is not sufficient to judge the overall performance of a CNN. When it comes to multi-class classification, it is necessary to consider how the classifications are distributed. The confusion matrix is one way of comparing the true or correct labels to the predicted labels. As is conventional for machine learning, rows correspond to the actual, true classes, while columns correspond to the predicted classes (i.e. the label returned by the neural network). With this convention, the cell (E, S0) refers to true ellipticals classified by the network as lenticulars. The confusion matrix of a perfectly accurate CNN is an identity matrix.

Figures \ref{fig:confmat3way} and \ref{fig:confmat4way} show confusion matrices for both the 3-way and 4-way classifications respectively, using the best-performing C2 neural network architecture, and trained on $100 \times 100$ resolution images. In the case of the 3-way classification, spirals are most accurately classified (93\%), following by ellipticals (80\%) and S0 (60\%). S0s are almost equally misclassified as E or Sp (18\% and 22\% respectively). Almost all the misclassified E were misclassified as S0, with just 1\% misclassified as spirals. Similarly, just 6\% of true spirals were misclassified as S0. Thus it is clear that ellipticals and spirals are the easiest morphological classes to distinguish between, while S0s are considerably more difficult.

In the 4-way classification case (Figure \ref{fig:confmat4way}), we find that both ellipticals and spirals have strong prediction accuracies (87\% and 89\% respectively). The accuracy of classifying S0s is unchanged, while the fourth irregular/miscellaneous class has a very poor accuracy only slightly higher than random. It is likely that the low accuracy of the irregular class is a contributing factor to the accuracy and loss fluctuations in Figure \ref{fig:traintest}. The majority of irregular samples are misclassified as spirals (50\%) or S0s (21\%). This is likely due to a combination of low-population categorical bias (there are only 388 irregular/misc samples out of 14034 total samples), and the fact that many miscellaneous samples show evidence of mergers, interaction and/or fine structure that may be interpreted as spiral or S0-like. This will be further analysed and discussed in Section \S 4, where we discuss hierarchical classification. 

Another means of assessing the performance of CNNs is to consider the overall F1-scores for each category. The F1-score is the harmonic mean of the precision and recall. In the context of machine learning, the precision of a neural network is defined as
\begin{equation*}
\text{Precision } = \frac{\text{TP}}{\text{TP}+\text{FP}}
\end{equation*}
where TP and FP denote true positives and false positives respectively. Precision, in this context, is equivalent to the positive predicative value or PPV. Recall is defined as
\begin{equation*}
\text{Recall } = \frac{\text{TP}}{\text{TP}+\text{FN}}
\end{equation*}
where FN denotes false negatives. It is equivalent to the sensitivity or true positive rate, as it measures the fraction of correctly identified positives. The F1-score is thus simply
\begin{equation*}
\text{F1 } = 2\times \frac{\text{Precision } \times \text{ Recall}}{\text{Precision } + \text{ Recall}}
\end{equation*}
and has a value between 0 and 1. The higher the F1-score, the better. Tables \ref{tab:f1scores} shows the precision, recall and F1-scores for our 3-way and 4-way classifications respectively. In general, spirals and ellipticals perform best. Although the precision is quite high for the irregular/miscellaneous samples, the recall is only slightly higher than random. This is indicative of the challenge in accurately classifying low populations of irregulars.

\begin{table}
\centering
\caption{Precision, recall and F1 scores for the best 3-way and 4-way classifications with our C2 network}
\label{tab:f1scores}
\begin{tabular}{|c|c|c|c|}
\hline 
Class \ & Precision & Recall & F1-Score \\ 
\hline 
E & 0.79 & 0.80 & 0.79 \\ 
S0 & 0.65 & 0.60 & 0.62 \\ 
Sp & 0.91 & 0.93 & 0.92 \\ 
\hline 
Weighted Avg & 0.82 & 0.83 & 0.83 \\
\hline
\hline
E & 0.75 & 0.87 & 0.81 \\
S0 & 0.61 & 0.60 & 0.60 \\
Sp & 0.91 & 0.89 & 0.90 \\
Irr+M & 0.48 & 0.27 & 0.35 \\
\hline
Weighted Avg & 0.80 & 0.80 & 0.80 \\
\hline
\end{tabular} 
\end{table}

\begin{figure}
\centering
\includegraphics[scale=0.68]{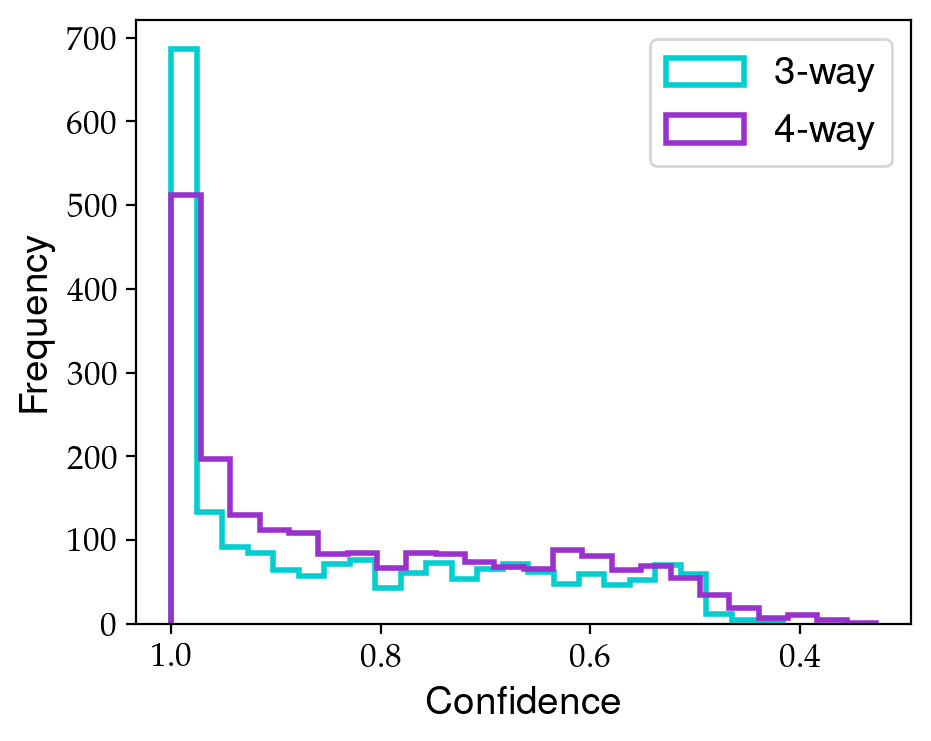}
\caption{Distribution of the confidences for each classified sample for both the 3-way (blue) and 4-way (purple) CNNs}
\label{fig:confidence}
\end{figure}

It is also useful to analyse the distribution of the confidence of each sample; that is, the probability that it belongs to the predicted class. As we have used a softmax activation function for the output layer, the activation values all sum to 1 across all categories. The predicted class is simply the category with the largest probability, and the value of that largest probability is known as the confidence. Figure \ref{fig:confidence} shows the distribution of these confidences for all samples in both the 3-way and 4-way classifications. Most samples are classified with a high degree of confidence, however the fact that the distribution is smooth at high to mid probabilities indicates that the CNN has not been unduly overfitted. In a highly overfitted CNN, the confidences would all cluster around 1 with little to no tail.

\section{Discussion}

\subsection{Classification Accuracies}

In general, our binary classification accuracies (Figure \ref{fig:hier-acc}) are comparable to those of other studies. Our accuracy of 98\% for ellipticals vs. spirals is in line with other studies \citep{DomingeuzSanchez2018, Barchi2020}, and also compares well with other early vs. late-type classifiers \citep{deDiego2020}. The key issue when it comes to training a binary classifier CNN is that there are an unequal number of samples for each class. This is most prominent with the irregular/miscellaneous samples, a mere 2.8\% of the total sample. With such a disparity, it is necessary to choose a subset of the overrepresented category in order to equalise the number of samples in each category, and hence reduce categorical bias. However, given the smaller overall dataset, there is a higher risk of overfitting, which the data augmentation is able to partially ameliorate.

Our best, direct 3-way classification accuracy of 84\% is also consistent with recent studies \citep{Barchi2020}, thought we note few studies have attempted to classify specifically between ellipticals, lenticulars and spirals, so a direct comparison is lacking. What is considerably promising is that our accuracy was achieved with a network trained only on single-band imagery, while the results of other studies are achieved with full colour images. This is especially advantageous when it comes to high-redshift surveys, or for galaxies imaged in only a single band. Figure \ref{fig:confmat3way} shows that spirals and ellipticals are most easiest to identify, while true lenticulars are spread between all three categories. It is important to note that there is a greater proportion of true ellipticals misclassified as S0 compared to true spirals. This is reflective of the lower E vs. S0 binary classification accuracy in Figure \ref{fig:hier-acc}. Similarly, given the comparatively higher S0 vs. Sp accuracy, we expect less spirals to be misclassified as S0.

Our direct 4-way classification accuracy of around 81\% is hampered by the difficulty in accurately classifying the irregular and miscellaneous samples. As seen in Figure \ref{fig:confmat4way}, most of the irregulars are classified as spirals, with one fifth also classified as S0s. The S0 accuracy is unchanged, while both ellipticals and spirals have slightly lower accuracies. The irregular class, which includes miscellaneous samples, contains myriad non-standard morphologies, and is particularly hampered by its low relative population. These factors all combine to affect the classification accuracies. We will discuss more about the impact of morphology in \S4.4.

\subsection{Physical Confusion Matrices}

\begin{figure*}
\centering
\includegraphics[scale=0.52]{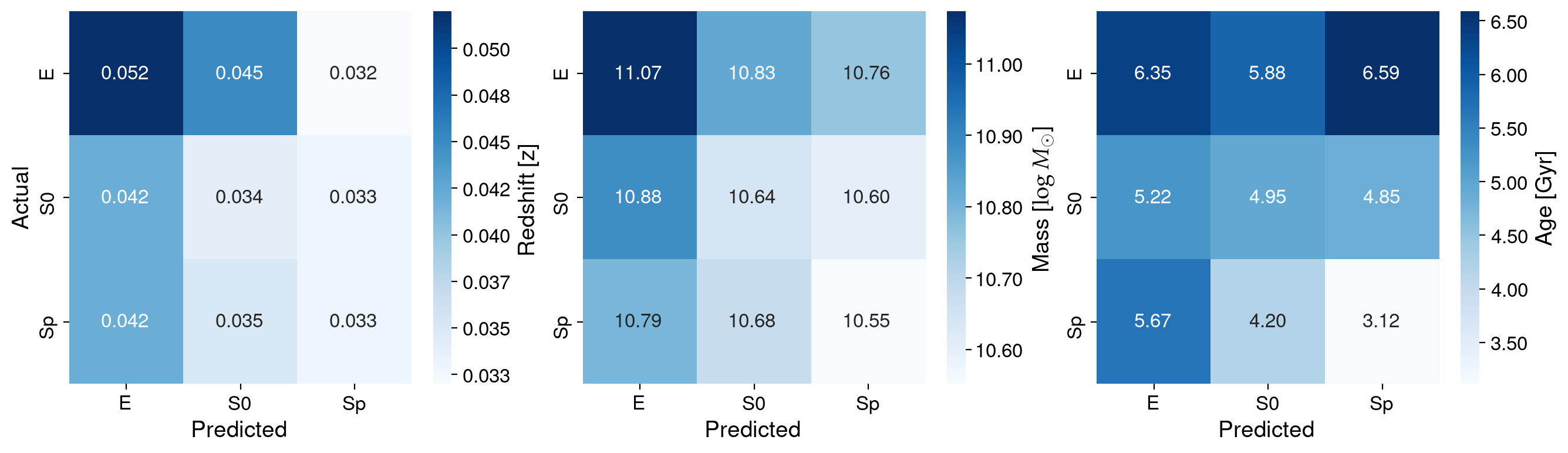}\\
\includegraphics[scale=0.52]{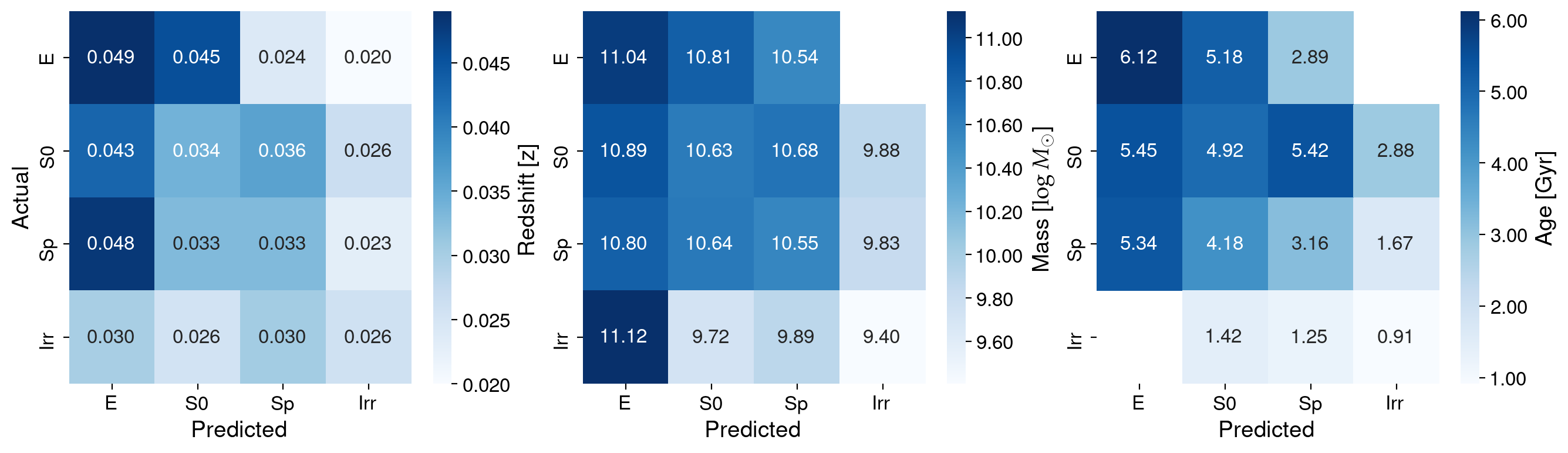}
\caption{Matrix plots, similar to the confusion matrices in Figures  \ref{fig:confmat3way} and \ref{fig:confmat4way}, where instead each cell shows the median redshift (left), stellar mass (center) and age (right) for each of the samples in the 3-way (top) and 4-way (bottom) CNN classifications of Figures \ref{fig:confmat3way} and \ref{fig:confmat4way} respectively. As an example, the cell (E, S0) in the top-left most panel shows the median redshift of all true ellipticals that were misclassified as lenticulars.}
\label{fig:physconf}
\end{figure*}

One way to glean insights into the nature of the CNN classifications is to consider the physical properties of each sample based on the cell that they occupy in the confusion matrix. These ``physical confusion matrices'' are shown in Figure \ref{fig:physconf} where we are calculating the median age, redshift and stellar mass for each sample based on their classification, e.g. median age of all true ellipticals classified as E, S0, Sp, all true spirals classified as E, S0, Sp, etc. All values are from the NA10 tables, with the values for physical quantities based on the MPA/JHU catalogues. As expected, true ellipticals have higher redshifts, masses and are older than their spiral counterparts. While this is expected given the general pathway of morphological evolution \citep{Conselice2014}, this is also reflective of the sample biases at higher redshifts where bright, compact objects like elliptical galaxies are more easily detected.

Given the values for the confused samples, there is evidence that even the misclassifications are physically meaningful. In the case of redshift, true ellipticals misclassified as S0s have (1) a lower median redshift than correctly classified ellipticals, and (2) a higher median redshift than correctly classified true S0s. Similarly, the median redshift is higher for both true S0s classified as ellipticals and true spirals classified as ellipticals. These trends are also observed for the other two properties: true ellipticals misclassified as S0 are less massive and younger than correctly classified true ellipticals, and are also more massive and older than correctly classified true S0s. These trends are also seen in the 4-way physical confusion matrices for all but the irregular class, where the low number of samples makes it difficult to establish any trends. Note that empty cells in some of the four-way confusion matrices are where no physical data is available for the sample(s) present. Importantly, the S0 and spiral samples misclassified as ellipticals are more distant, more massive and older than true S0s and spirals. Each of these characteristics are indeed to be expected of an elliptical galaxy. Thus this demonstrates that the CNN has learned to identify features that discriminate between the different morphological types. If the misclassifications were random then we would not see such a trend across multiple physical properties.

\begin{figure*}
\centering
\includegraphics[scale=0.57]{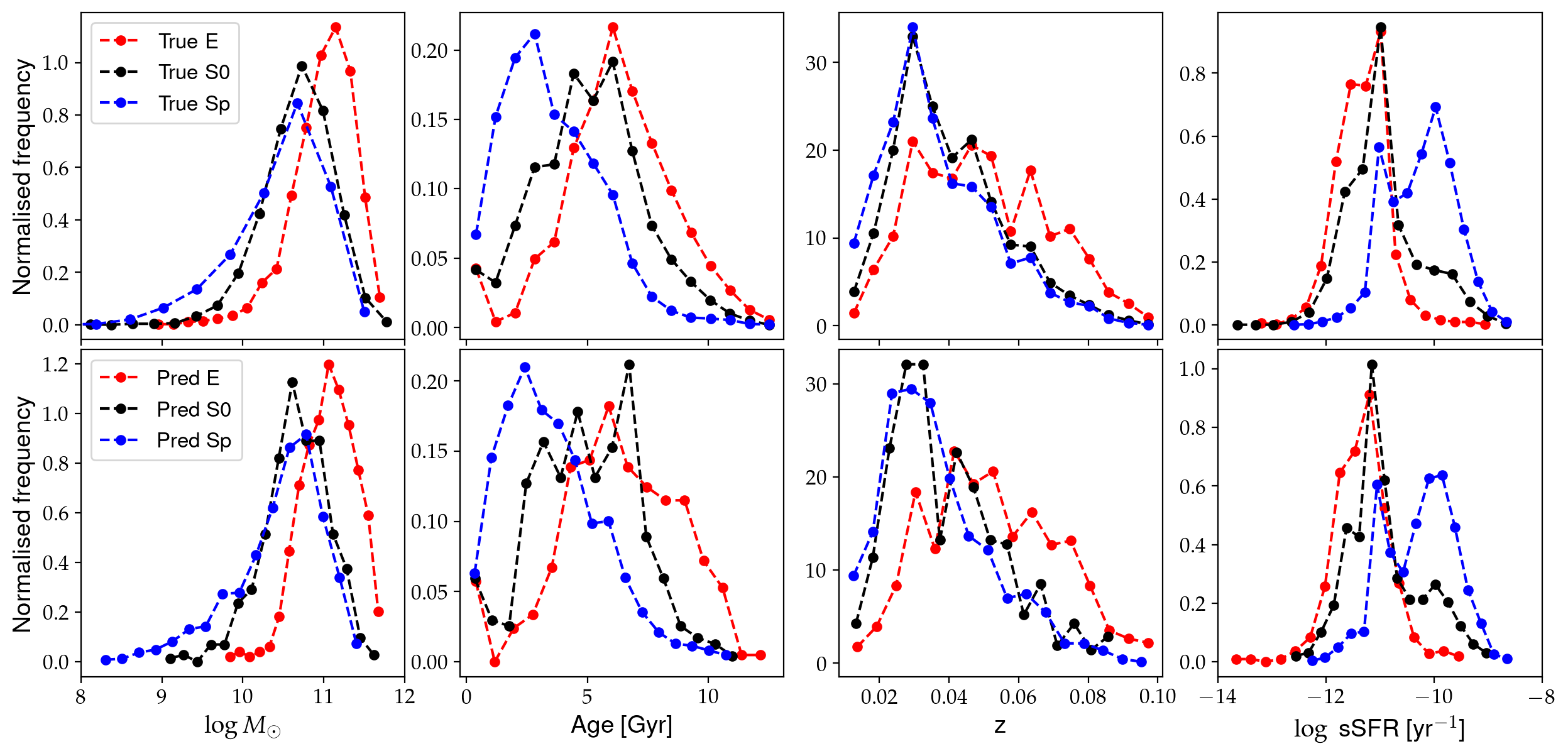}
\caption{Plots of the normalised frequency for the E, S0 and Sp morphological classes as functions of stellar mass $M_\odot$, age (in Gyr), redshift $z$ and specific star formation rate $\log r_{\text{SFR}}$. The top row is for all true morphological samples in the complete dataset, while the bottom panels are for the predicted morphologies of the samples in the testing set as classified by the 3-class CNN as either E (red), S0 (black) or Sp (blue).}
\label{fig:hist-3way}
\end{figure*}

\begin{figure*}
\centering
\includegraphics[scale=0.57]{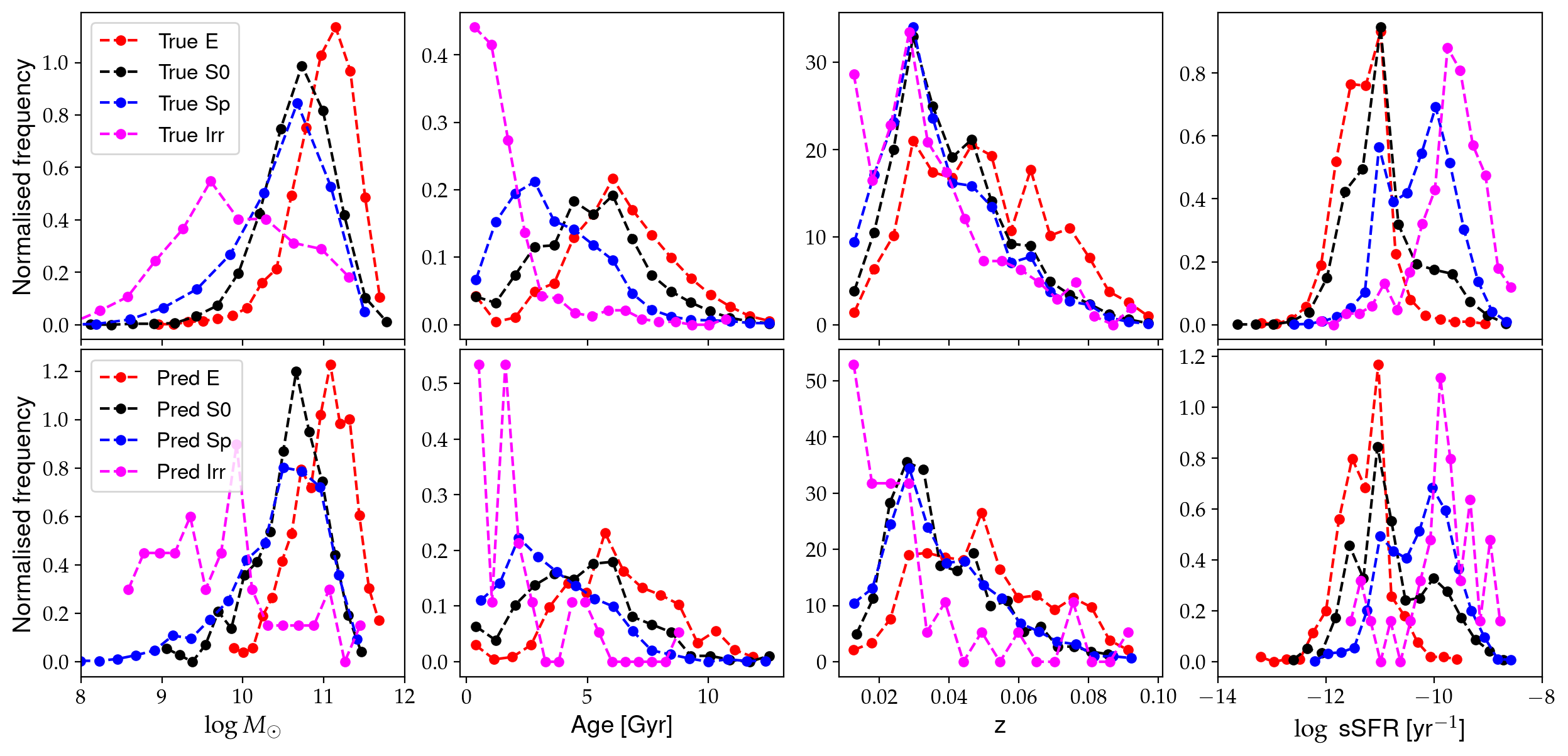}
\caption{Similar plot as in Figure \ref{fig:hist-3way}, albeit with the addition of the fourth morphological class. The top row plots the true morphologies of the complete dataset, while the bottom plots predicted morphologies for samples in the testing set as classified by the 4-class CNN as either E (red), S0 (black), Sp (blue) or Irr+Misc (magenta).}
\label{fig:hist-4way}
\end{figure*}

Another way to visualise these trends is by comparing the distribution of the physical properties for the predicted morphological types with that of the true morphological types of the entire dataset. This is to see whether the CNN classifications are physically reasonable, as Figure \ref{fig:physconf} contends. Figures \ref{fig:hist-3way} and \ref{fig:hist-4way} show plots of the (normalised) density distribution of the stellar mass, age, redshift and specific star formation rate for both the true morphological types of all samples (top panels), and predicted morphological types of all the test set samples (bottom panels). Both the distributions are well matched, suggesting that the predicted morphologies share the same general distributions of physical properties as the actual morphologies. As expected, ellipticals are more massive, older, more distant and are largely quenched. Likewise, the distribution of the physical properties of predicted spiral galaxies is consistent with a quintessentially low-mass, young, low-redshift, actively star-forming population. The particular case of the irregular / miscellaneous samples in Figure \ref{fig:hist-4way} is more problematic given the low population and its broad categorisation. This categorisation, which includes interacting galaxies / mergers, is reflected in the wider distribution in stellar masses. However, the distribution of irregulars does peak at the youngest overall age, in both true and predicted morphology.

\subsection{Hierarchical Classification}

\begin{figure}
\centering
\includegraphics[scale=0.48]{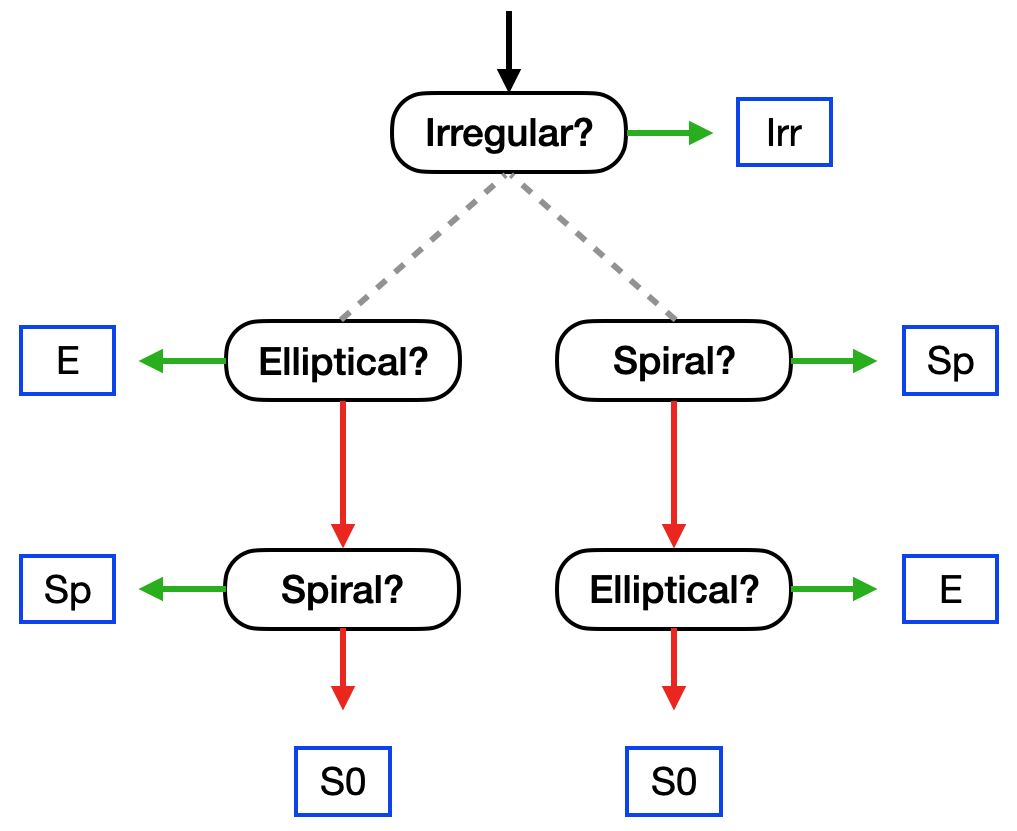}
\caption{Diagram showing the two different paths in our hierarchical classification tree. In the case of 3-way classification, the first step is ignored. To classify between E, S0 and Sp, there are two routes depending on whether to first look for ellipticals or look for spirals. The subsequent step is then to discriminate between the other two categories.}
\label{fig:hiertree}
\end{figure}

\begin{figure*}
\centering
\includegraphics[scale=0.68]{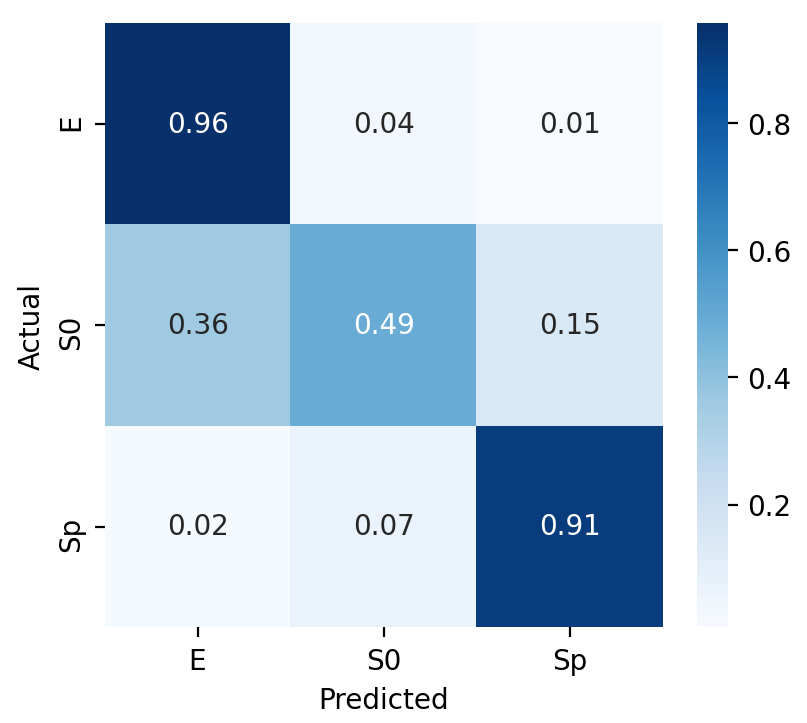}\hspace{3mm}
\includegraphics[scale=0.68]{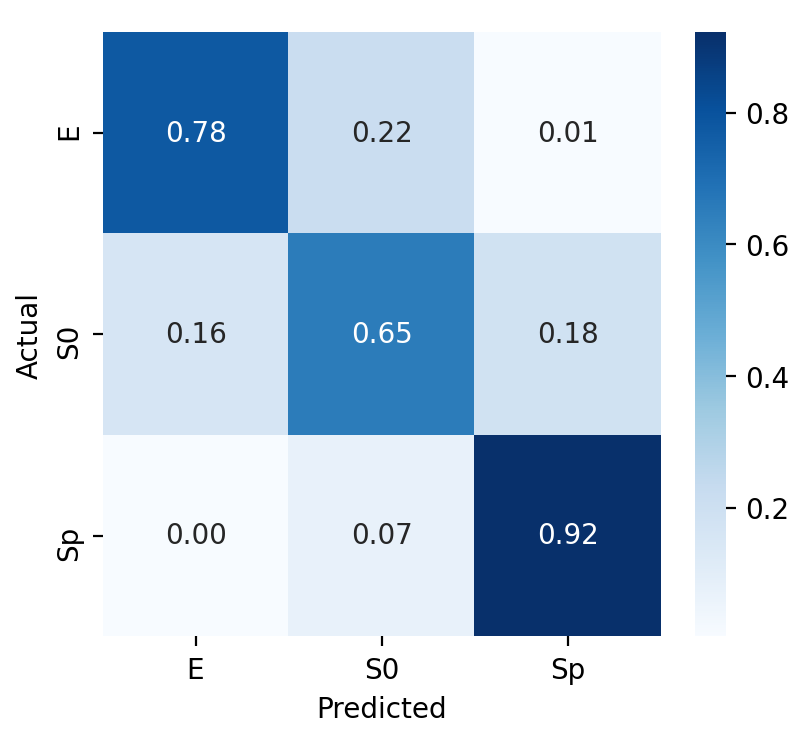}
\caption{Confusion matrices for the 3-way hierarchical classification of ellipticals (E), lenticulars (S0) and spirals (Sp) for the E-first pathway (left) and Sp-first pathway (right) (see Figure \ref{fig:hiertree} for an overview of these pathways).}
\label{fig:3hconf}
\end{figure*}

%\begin{figure*}
%\centering
%\includegraphics[scale=0.65]{figures/4h_conf_efirst}\hspace{5mm}
%\includegraphics[scale=0.65]{figures/4h_conf_spfirst}
%\caption{Confusion matrices for the 4-way hierarchical classification (E, S0, Sp, Irr) for the E-first pathway (left) and Sp-first pathway (right).}
%\label{fig:4hconf}
%\end{figure*}

%\begin{figure}
%\centering
%\includegraphics[scale=0.65]{figures/4h_conf_regthen3way}
%\caption{Confusion matrix for the 4-way hierarchical classification (E, S0, Sp, Irr). Here the first step involves classifying between irregulars and regulars (Irr vs. Reg, see Figure \ref{fig:hiertree}), and the second step is an application of our direct 3-way CNN. This confusion matrix is the best out of all the hierarchical 4-way classifications.}
%\label{fig:4hconf3way}
%\end{figure}

\begin{table}
\centering
\caption{Table of classification accuracies for all five hierarchical classifiers.}
\label{tab:hmetrics}
\begin{tabular}{|c|c|c|}
\hline 
Path & Best 3-way Accuracy & Best 4-way Accuracy \\ 
\hline 
E-first & 82\% & 63\% \\ 
\hline 
Sp-first & 83\% & 63\% \\ 
\hline 
Reg+3-way & N/A & 65\% \\ 
\hline 
\end{tabular} 
\end{table}

In this work, we have primarily trained CNNs to directly classify galaxy samples directly as either one of three or four classes. It is thus useful to compare this direct comparison with an indirect classification method involving hierarchical classification. This indirect method was also explored in an attempt to obtain better classification accuracies for the irregulars. Here, we use multiple CNNs to construct a binary decision tree, with which to classify samples. This approach has been used in many previous works, fundamentally as part of the Galaxy Zoo \citep{Lintott2008}, as well as in other studies in astronomy involving machine learning \citep{HuertasCompany2010, DomingeuzSanchez2018, HuertasCompany2019}. Here we simply process each sample in the test set by passing it through a series of binary CNNs (see Figure \ref{fig:hier-acc} for the classification accuracies of each). We can perform the 3-class classification using two different hierarchical classifiers; first, classify between E and S0+Sp, then, given the sample is not an elliptical, between S0 and Sp. The second route is to first classify between E+S0 and Sp, and then, given the sample is not a spiral, between E and S0. In the case of the 4-class classification, the first binary CNN is simply E+S0+Sp (Regular) vs. Irr+Misc. Figure \ref{fig:hiertree} illustrates these two different pathways. We will refer to the pathway that commences with E vs. S0+Sp as E-first, and E+S0 vs. Sp as Sp-first. We also classify between the four classes first by applying the Regular vs. Irr+Misc binary classifier, and then applying our direct 3-way CNN (thus two overall steps instead of three). We refer to this as our Reg+3-way classifier. Table \ref{tab:hmetrics} lists the best overall classification accuracies for each of the five hierarchical classifiers.

Figure \ref{fig:3hconf} shows the confusion matrices for the hierarchical 3-way classification with both the E-first and Sp-first pathways. Both have similar overall accuracies, but the Sp-first path best matches the direct 3-way CNN (Figure \ref{fig:confmat3way}) with a better 65\% S0 accuracy and comparable E and Sp accuracies. The E-first path, while excellent at classifying ellipticals, has a much poorer S0 accuracy, with over a third of true S0s misclassified as ellipticals. This is despite the fact that both the E vs. S0+Sp and E+S0 vs. Sp binary classifiers have similar >90\% accuracies. The disparity is reflective of a more deeper issue: the fact that ellipticals and S0s are inherently much harder to disentangle, compared to distinguishing between spirals and S0s (compare the 82\% and 88\% binary classification accuracies in Figure \ref{fig:hier-acc}). This may seem counter-intuitive; after all, if Sp are able to be accurately picked up, then the remaining E+S0 samples should have a worse accuracy. The key point is that this is dependent on the percentage of true spirals. Spirals are able to be differentiated between E and S0 with 98\% and 88\% accuracies respectively, so fewer false positives will progress to the next stage in the hierarchical tree. However, because E vs. S0 has a lower accuracy, more false positive ellipticals (in particular, true S0s falsely classified as E) will emerge from the first E vs. S0+Sp stage. Hence the E-first path results in a greater spread of samples, while the Sp-first path is more accurate at picking up true S0s. For both the Sp-first pathway in Figure \ref{fig:3hconf} and the direct 3-way CNN in Figure \ref{fig:confmat3way}, more true ellipticals are misclassified as S0s compared to true spirals.

\begin{table}
\centering
\caption{Table of per-class classification accuracies for the three 4-way hierarchical classifiers (E-first, Sp-first and Reg+3-way).}
\label{tab:h4wayc}
\begin{tabular}{|c|c|c|c|c|}
\hline 
CNN & E & S0 & Sp & Irr \\ 
\hline 
E-first & 91\% & 33\% & 63\% & 90\% \\ 
\hline 
Sp-first & 76\% & 45\% & 64\% & 90\% \\ 
\hline 
Reg+3-way & 80\% & 46\% & 65\% & 90\% \\ 
\hline 
\end{tabular} 
\end{table}

Table \ref{tab:h4wayc} shows the per-class classification accuracies for each of our 4-way hierarchical CNNs. Note from Table \ref{tab:hmetrics} that the E-first and Sp-first had similar overall accuracies (63\%), with a slightly higher 65\% for the Reg+3-way CNN. Adding an explicit step to classify irregulars has dramatically improved the classification accuracy from 27\% to 90\%, but this has come at the severe cost of reduced accuracies for other classes. This suggests that the more steps there are in the pathway, the harder it is to classify the classes in the later steps. Although the E-first 4-way CNN can classify ellipticals with a 91\% accuracy, the S0 accuracy is the lowest out of the three (33\%). The overall Sp and Irr classification accuracies are similar across all three classifiers, with the Irr accuracy unchanged. As with the 3-way hierarchical classifiers, the Sp-first approach provides a better spread of accuracies across all the categories. The Reg+3-way CNN offers a further marginal improvement.

These results show that indirect, hierarchical classification can offer comparable accuracies to direct 3-way classification, but performs poorly when compared to direct 4-way classification. Although the direct CNNs outperform the hierarchical classifiers, the 3-way Sp-first classifier offers a 5\% improvement in S0 classification accuracy.

\subsection{Galaxy Morphology}

\begin{figure*}
\centering
\includegraphics[scale=0.46]{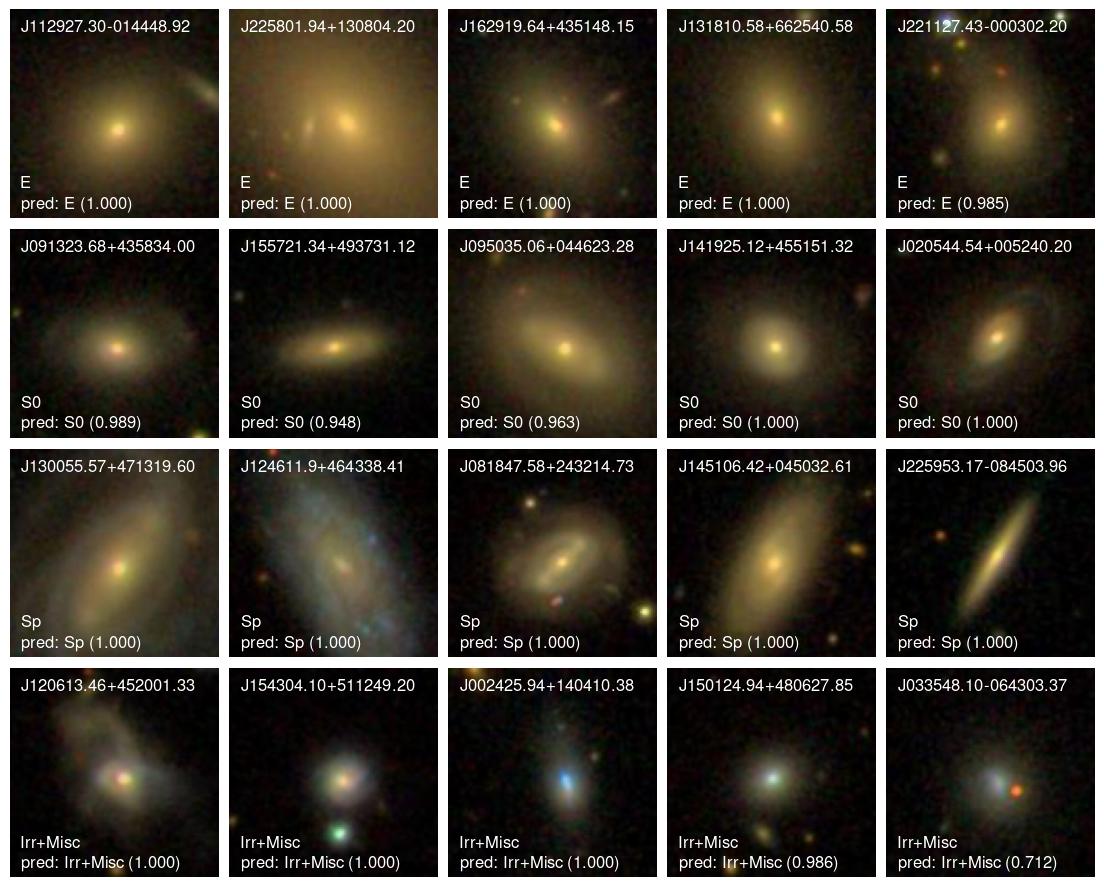}
\caption{Randomly selected galaxy samples from each of the 4 morphological classes that were correctly classified by the CNN. Each image is annotated, from top to bottom, with its SDSS JID, actual morphological class, and predicted morphological class, with the CNN prediction confidence given in parentheses.}
\label{fig:ex-correct}
\end{figure*}

\begin{figure*}
\centering
\includegraphics[scale=0.38]{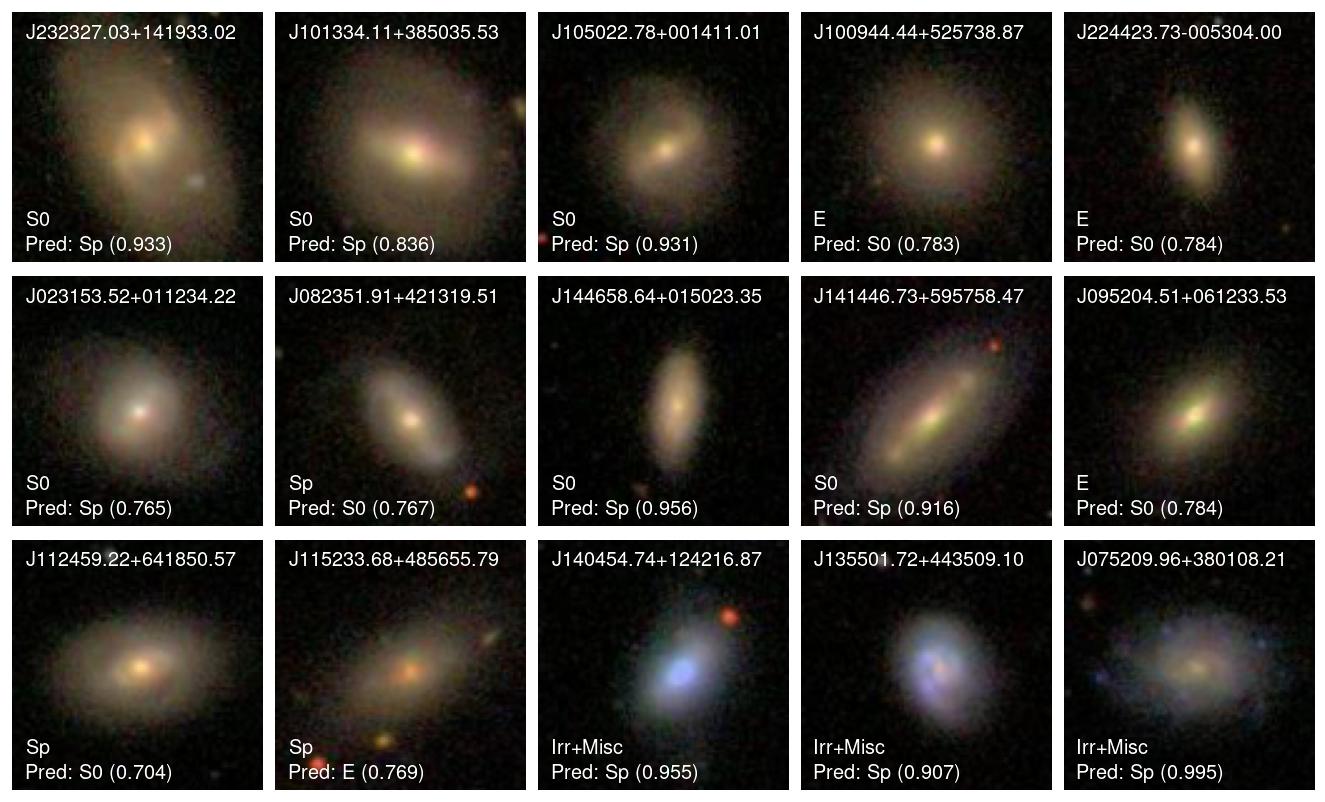}
\caption{Randomly selected galaxy samples from each of the 4 morphological classes that were incorrectly classified by the CNN with a relatively high confidence (>70\%). Each image is annotated with similar text as in Figure \ref{fig:ex-correct}}
\label{fig:ex-incorrect}
\end{figure*}

Galaxies exhibit an array of fine morphological structure, from spiral arms, dust lanes and stellar bars, to tidal tails and/or tidal interaction \citep{Conselice2014}. The power of CNNs lie in their ability to extract features in images. As can be seen in the example in Figure \ref{fig:cnn-vis}, the convolutional layers are able to outline, trace and detect the presence of spiral arms. The process of visual classification relies on being able to detect and categorise such morphological features \citep{Buta2013}. In general, S0 galaxies are difficult to visually distinguish \citep{Blanton2009}, in part since they exhibit morphological characteristics that, insofar as solely examining an image is concerned, are near identical to ellipticals. S0s can also display prominent dust lanes, rings and large disc components, which make them visually similar to spirals, especially in the edge-on case. It is thus expected that the overall CNN classification accuracies are lower for S0s. Although the networks were trained on single-band images (g-band), for the purpose of illustrations all examples in the figures of this section are the colour image cut-outs taken directly from SDSS DR4.

Figure \ref{fig:ex-correct} shows a random selection of galaxies, from each morphological class, that were correctly identified by the CNN. Most of these samples were classified with a near-certain confidence (>0.99). Although the ellipticals and S0s in Figure \ref{fig:ex-correct} look visually similar, the CNN is able to distinguish between the two types, possibly by being able to detect the more prominent bulge and/or evidence of disc-like structure as seen in the S0s. Figure \ref{fig:ex-incorrect} shows a selection of fifteen randomly chosen samples that were misclassified with a relatively high confidence, at least >70\%. The first three samples (starting from the top-left and moving across to the right) are all S0s that were misclassified as spirals. All three exhibit fine morphological structure, such as a prolate bulges and stellar bars, and there are hints of faint spiral arms in each. Whether such arms are truly well established enough to consider the galaxy a spiral is debatable, but these examples show that the CNN is able to both detect their presence and classify accordingly. J100944.44+525738.87, an elliptical classified as an S0, demonstrates the difficulty in distinguishing between these two types by image alone. There are also examples of clear confusion; the first two samples in the middle row, J023153.52+011234.22 and J082351.91+421319.51, are both barred galaxies. They have a dataset classification of S0 and Sp respectively, but have been classified by our CNN as opposites (Sp and S0, respectively). The CNN has also misclassified all the irregular samples in Figure \ref{fig:ex-incorrect} as spirals. Since the network is only trained on g-band images, the prominent blue colour of the irregulars that enables the naked eye to easily distinguish that morphological type is instead lost to the CNN. One other interesting example is that of J112459.22+641850.57, a spiral galaxy with extremely faint arms that arguably looks like a stellar ring, which the CNN has classified as an S0.

\begin{figure*}
\centering
\includegraphics[scale=0.28]{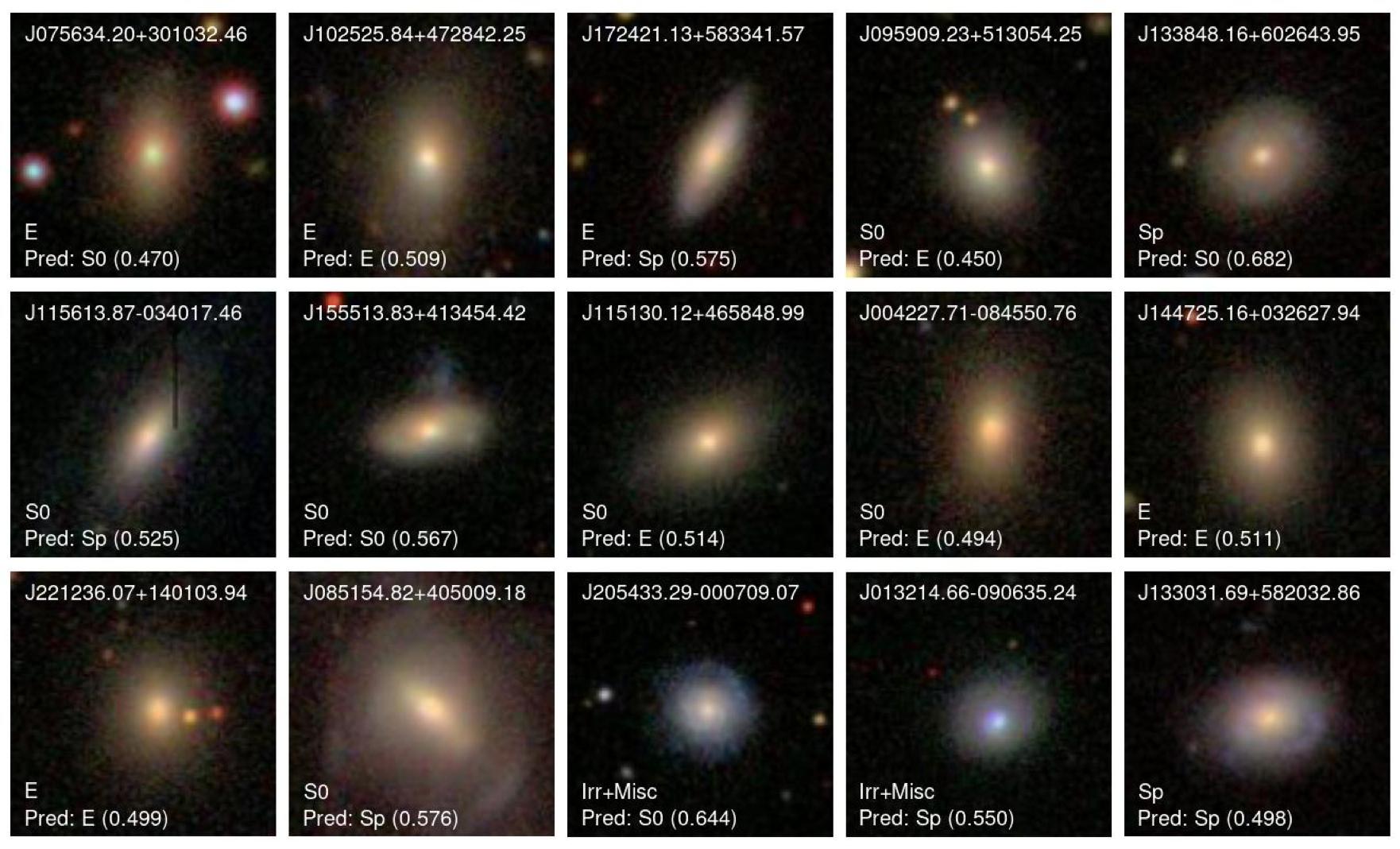}
\caption{Randomly selected galaxy samples from each of the 4 morphological classes that were classified by the CNN with a relatively low confidence (<70\%, mostly <60\%). Each image is annotated with similar text as in Figure \ref{fig:ex-correct}}
\label{fig:ex-lowconf}
\end{figure*}

\begin{figure*}
\centering
\includegraphics[scale=0.55]{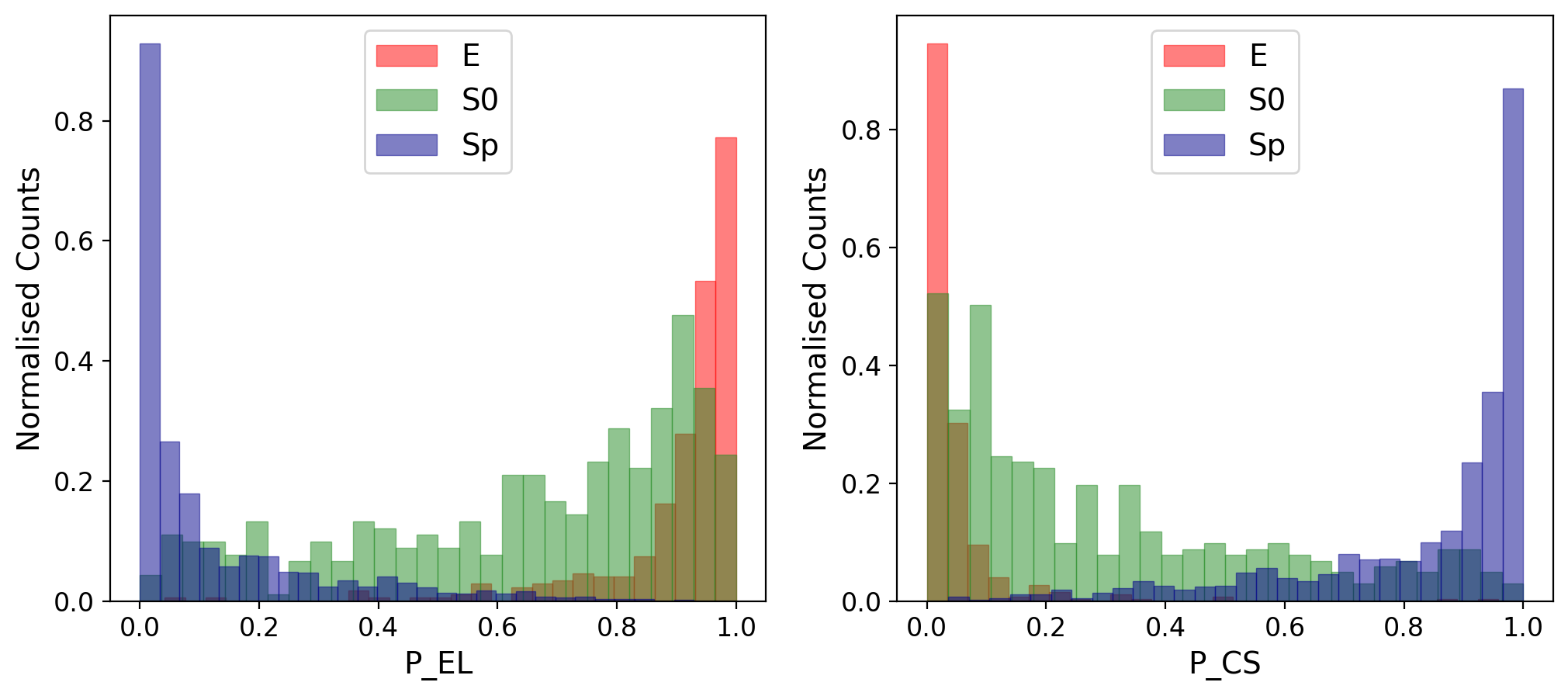}
\caption{Distribution of all samples in our main NA10 test set, classified by the CNN as either elliptical (red), S0 (green) or spiral (blue), plotted with respect to the Galaxy Zoo probability of being an elliptical.}
\label{fig:pel-pcs}
\end{figure*}

Figure \ref{fig:ex-lowconf} shows a random selection of samples that were classified with confidences less than 70\%. These examples include samples that were both correctly and incorrectly classified, and can help reveal where most of the confusion is based. In general, compared to Figures \ref{fig:ex-correct} and Figures \ref{fig:ex-incorrect}, there is far less variety in large-scale morphology, and instead distinctions must lie in finer-scale morphologies as well as foreground/background features. The sample images containing foreground objects in Figure \ref{fig:ex-lowconf} tend to result in lower confidences, most spectacularly with the first example, J075634.20+301032.46. The last two examples in the second row, J004227.71-084550.76 and J144725.16+032627.94, both near visually identical, have dataset classifications of S0 and E respectively, but both have been classified by the CNN as ellipticals with 49\% and 51\% confidence respectively. What is also important to note is the uncertainty that surrounds the classification of edge-on galaxies. In particular, the middle image in the top row of Figure \ref{fig:ex-lowconf} and the first image in the second row (J172421.13+583341.57 and J115613.87+034017.46) are both edge on galaxies, and they have both been classified as spiral galaxies. Indeed, since many spiral galaxies have prominent discs, from pure population statistics alone it is more likely that an edge-on galaxy is a spiral, rather than an elliptical. This is also problematic for S0 galaxies, many of which also have a prominent disk and are difficult to distinguish from spirals when viewed edge on. This is where other analytical techniques are needed to definitively tell the two apart (e.g \citealt{Laurikainen2005}).

A further way to check whether our CNN classifications are morphologically sensible is to compare our predicted morphologies with the probabilistic Galaxy Zoo morphologies for all the NA10 samples in our test set. This is important, as it allows us to compare the CNN classification confidences with the Galaxy Zoo probabilities that a sample is an elliptical or spiral. Here, we performed a cross-match and obtained the Galaxy Zoo probabilities P\_EL, the probability of being an elliptical, and P\_CS, the combined probability of being a spiral \citep{Lintott2011}. Figure \ref{fig:pel-pcs} gives the normalised distributions of the samples in our NA10 test set that were classified as ellipticals (red), S0 (green) and spirals (blue), in terms of Galaxy Zoo probability of being an elliptical or spiral. As expected, all NA10 predicted spirals have a very low probability of being an elliptical, and very high probability of being a spiral. Similarly, all NA10 predicted ellipticals have high probabilities of being an elliptical, and low probabilities being a spiral. The S0s have significantly more spread, albeit are grouped more towards being an elliptical. This much broader distribution is consistent with the difficulty in definitively classifying S0s, as implied by the low confidence examples in Figure \ref{fig:ex-lowconf}. This also further justifies why so many ellipticals were classified as S0s in the 3-way and 4-way classifications in Table \ref{tab:gzacc}.

\begin{table}
\centering
\caption{Classification spreads and accuracies for the binary, direct 3-way, hierarchical 3-way and direct 3-way CNNs when applied to samples of 1,000 ellipticals and 1,000 spirals.}
\label{tab:gzacc}
\includegraphics[scale=0.38]{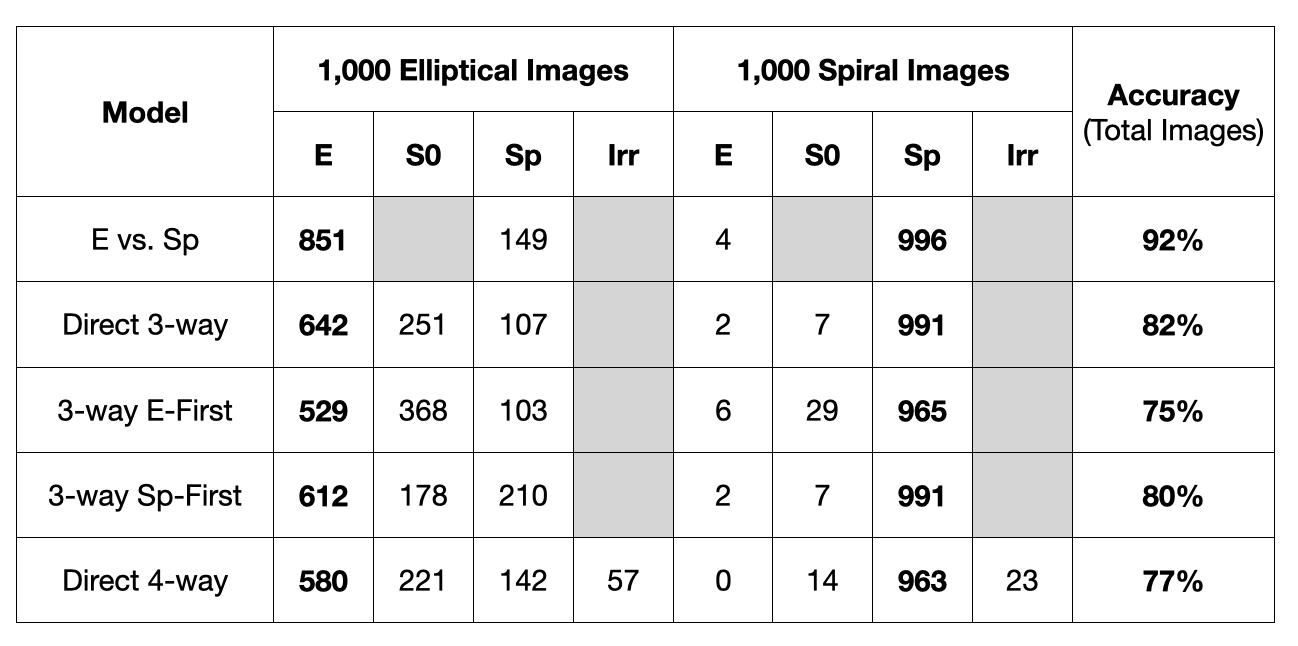}
\end{table}

\subsection{Galaxy Zoo Sample Test}

\begin{figure*}
\centering
\includegraphics[scale=0.28]{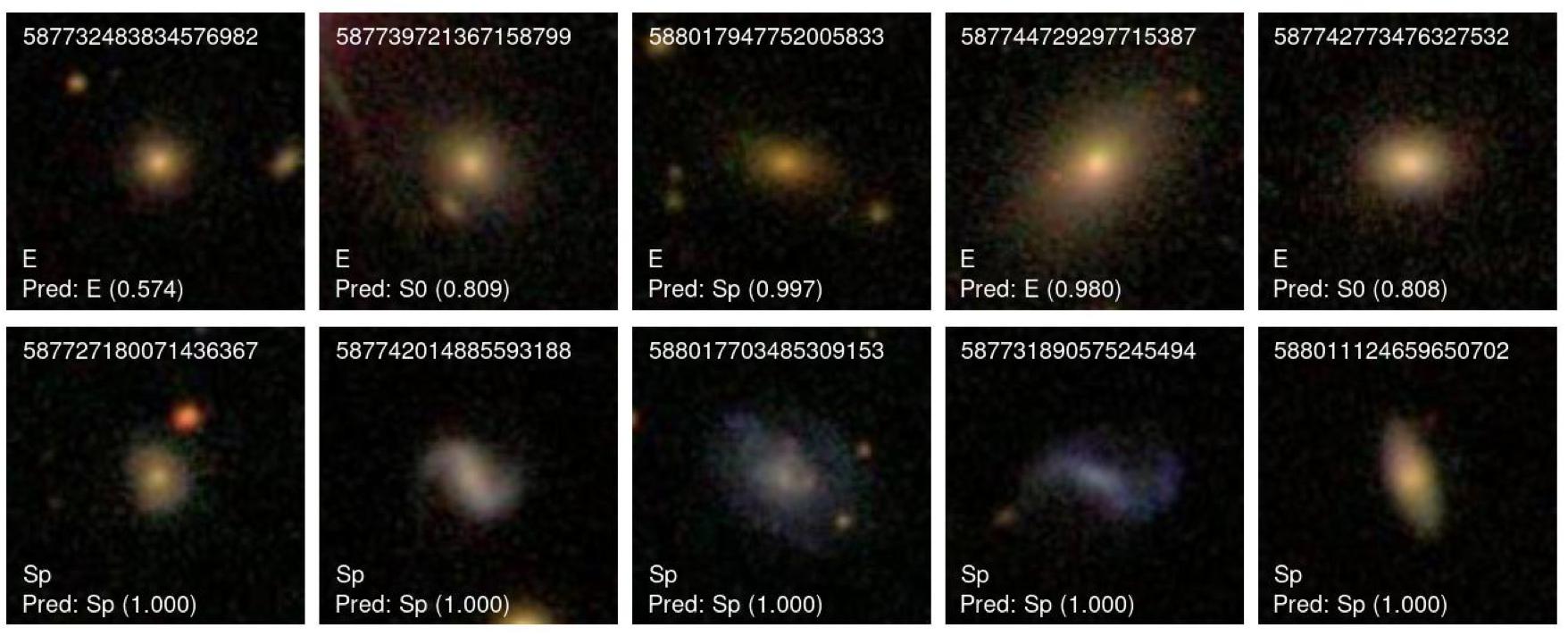}
\caption{Randomly selected elliptical and spiral galaxies from our Galaxy Zoo subsample. Images are annotated with their Galaxy Zoo OBJID at the top, along with the Galaxy Zoo class, followed by the CNN predicted class with the confidence of the prediction given in parentheses.}
\label{fig:ex-gz}
\end{figure*}

In addition to our tests with the NA10 dataset, we further apply five of our networks to a small yet unique sample of Galaxy Zoo images \citep{Lintott2008}. In particular, we chose a random selection of 1,000 ellipticals and 1,000 spiral galaxies from the Galaxy Zoo catalogue that were not included in the NA10 dataset. These images were classified via the collective contribution of citizen scientists as either elliptical, spiral or uncertain. Table \ref{tab:gzacc} gives the number of samples classified in each morphological class, as well as the overall accuracy, for each of the five networks: binary elliptical vs. spiral, direct 3-way CNN, the two 3-way hierarchical classifiers (E-first and Sp-first), and the direct 4-way CNN. All CNNs excelled at classifying spirals with very high accuracies. However, the performance for the ellipticals is not as spectacular, with 85\% for the binary CNN, dropping to 64\% and 58\% for the direct 3-way and 4-way CNNs respectively. There are two main reasons for this. Firstly, spirals have the highest per-class accuracy out of all the classes; this is consistent across all networks (see Figures \ref{fig:confmat3way}, \ref{fig:confmat4way}, \ref{fig:3hconf}, etc). Secondly, the Galaxy Zoo elliptical categorisation does not explicitly distinguish between elliptical and S0 galaxies. For the direct 3-way CNN, as seen in Table \ref{tab:gzacc}, out of the 1000 ``ellipticals'', 251 are classified as S0s. This is even worse in the E-first hierarchical classifier, with 368 classified as S0s. These problems become apparent for 3-way classification and above. However, throughout all the CNNs, there are some ellipticals that have been misclassified as spirals. It is likely that these are edge-on ellipticals. Figure \ref{fig:ex-gz} shows a random selection of these elliptical and spiral samples, along with their predicted classes and confidence values as classified by our direct 3-way CNN. All spirals are correctly classified with 100\% certainty, while the ellipticals generally have lower classification confidences.

\subsection{Efficiency and Limitations}

With the advent of large-scale structure surveys comes the need to process enormous amounts of data efficiently and accurately. This is crucial in the context of classifying galaxy morphologies, for which visual classification is intractable. The key strengths of automated classification techniques, such as our CNN approach, ultimately lie in their speed and ability to generalise. Although training a CNN can be a computationally expensive undertaking, the speed with which it can classify galaxies once trained is orders of magnitude greater than what could ever be possible with manual classification. Our C2 network processed the test set in under five seconds using a GTX 1080Ti GPU; for comparison, it took a group of classifiers at least five minutes to reach a consensus on the samples in Figure \ref{fig:ex-incorrect}. The ability to generalise is another key strength of CNNs as, by design, they offer a model-independent means of extracting relevant features from arbitrary input data. Trained CNNs can also be easily adapted to classify other datasets through the use of transfer learning (see \citealt{DomingeuzSanchez2019}).

Despite these advantages, there are some drawbacks to our current implementation. The use of single-band images necessary limits the amount of information available to the CNN. Although this work has achieved comparable overall accuracies to other studies using colour images, combining multiple bands will make it easier to distinguish between different morphologies, especially irregulars that tend to be bluer. Our current CNN approach relies solely on images, yet kinematic data is useful at distinguishing between morphological types, particularly between E and S0 \citep{Cappellari2007}. Studies based on simulations have successfully combined density and velocity maps for improved classification \citep{Shen2020}, and so the incorporation of kinematic data  is a promising area of future study.

\section{Conclusion}

In this work, we trained and tested several convolutional neural networks to classify galaxy morphologies. This work is among the first to explicitly classify ellipticals, S0 galaxies and spirals with a direct 3-class CNN, in addition to a fourth irregular class with a direct 4-class CNN. The main outcomes of this work are summarised as follows:

\begin{itemize}
\item We introduced a new CNN architecture that can classify between ellipticals, lenticulars and spirals with an accuracy of 84\%, and between E, S0, Sp and irregulars/miscellaneous with an accuracy of 81\%. These accuracies outperform the results of existing models at 3-way classification, and has equivalent accuracy for 4-way classification.
\item We demonstrated that a network using our new architecture has equivalent performance on single-band imagery compared to other studies that have utilised colour imagery. Although this is largely domain-specific, it gives confidence that CNNs can be applied to datasets that only contain information for a single-band. This is especially important for high-redshift surveys.
\item We also used our new architecture to train several binary classifiers to classify between the four chief morphological types. We found that our CNN is best able to distinguish between ellipticals and spirals with an accuracy of 98\%, while spirals and irregulars were the hardest to distinguish between, with an accuracy of only 78\%.
\item We showed that the direct 3-class CNN has a marginally better accuracy compared to an indirect, hierarchical classifier, but that the direct 4-class CNN is more accurate.
\item Through an analysis of the confusion matrices, and the physical properties of the misclassified samples, we have shown that even the misclassified samples are physically sensible. In particular, samples predicted by the CNN to be ellipticals (whether correctly or incorrectly) tended to have higher redshifts, higher stellar masses and were older in age. Similar trends were observed in the case of S0s misclassified as E (as compared to true S0s), and Sp classified as E (compared to true spirals). The distributions of the physical properties of the predicted morphologies also match well with the true morphologies.
\item Through inspecting several samples classified with low confidences, we found that ellipticals and S0s, often with a mostly featureless face, were harder to distinguish between. We found that many edge-on galaxies were often being associated with spiral galaxies, and this is likely due to the the effect of population statistics inherent in the overall dataset. We found that the low overall number of irregular samples also contributed to difficulties in the 4-way classification, with most irregular and miscellaneous samples classified as spirals, likely due to their myriad non-standard morphologies.
\end{itemize}

\section*{Acknowledgements}

We are thankful for the detailed and constructive feedback from the anonymous referee that helped to improve this paper. This research made use of the Keras deep learning API \citep{Chollet2015} and the TensorFlow machine learning library (see \url{https://www.tensorflow.org/}). We acknowledge ICRAR for providing the necessary compute resources to carry out this research via its \textit{Hyades} GPU cluster. MKC acknowledges the financial support of an Australian Government Research Training Program Scholarship at The University of Western Australia.

\section*{Data Availability}

The data underlying this article will be shared on reasonable request to the corresponding author.

\bibliographystyle{mnras}
\bibliography{mybib} % if your bibtex file is called example.bib

%\appendix

% Don't change these lines
\bsp	% typesetting comment
\label{lastpage}
\end{document}